\documentclass[%
 reprint,
 amsmath,amssymb,
 aps,prx
]{revtex4-2}

\usepackage{graphicx}
\usepackage{dcolumn}
\usepackage{bm}
\usepackage{xcolor}
\usepackage{physics}
\usepackage[caption=false]{subfig}
\usepackage{ragged2e}
\usepackage{mwe}
\usepackage{tcolorbox}
\usepackage{bbold}
\usepackage{tkz-euclide}
\usepackage{wasysym}
\usepackage[T1]{fontenc}
\usepackage{blindtext, rotating}
\DeclareCaptionJustification{justified}{\justifying}
\renewcommand{\vec}[1]{{\boldsymbol #1}}
\newcommand{\new}[1]{{\color{black} #1}}

\begin{document}

\preprint{APS/123-QED}

\title{Dynamics of visons and thermal Hall effect in perturbed Kitaev models}

\author{Aprem P. Joy}
\author{Achim Rosch}%
\affiliation{%
 Institute for Theoretical Physics, University of Cologne, Cologne, Germany
}%
\date{\today}

\begin{abstract}
A vison is an excitation of the Kitaev spin liquid which carries a $\mathbb Z_2$ gauge flux. While immobile in the pure Kitaev model, it becomes a dynamical degree of freedom in the presence of perturbations.
We study an isolated vison in the isotropic Kitaev model 
perturbed by a small external magnetic field $h$, an offdiagonal exchange interactions $\Gamma$ and a Heisenberg coupling $J$. In the ferromagnetic Kitaev model, the dressed vison obtains a dispersion linear in $\Gamma$ and $h$ and a fully universal low-$T$ mobility, $\mu=6 \hbar v_m^2/(k_B T)^{2}$,
where $v_m$ is the velocity of Majorana fermions. In contrast, in the antiferromagnetic Kitaev model interference effects \new{suppress} coherent 
propagation and an incoherent Majorana-assisted hopping  leads to a $T$-independent mobility.
The motion of a single vison due to Heisenberg interactions is strongly suppressed for both signs of the Kitaev coupling.
Vison bands in AFM Kitaev models can be topological and may lead to characteristic features in the thermal Hall effects in Kitaev materials.
\vspace{0.1\textwidth}
\end{abstract}

\maketitle

\section{\label{sec:intro}INTRODUCTION}

Gauge theories are central to our understanding of high-energy physics where they mediate interactions between fundamental particles. While in the standard model the existence of gauge symmetries is postulated, they `emerge' naturally in the description of certain strongly correlated solid-state systems. Such systems host fractional excitations with exotic quantum numbers. In this context, one of the best understood models is the honeycomb Kitaev model which hosts a spin liquid in its ground state \cite{Kitaev06}. In this two-dimensional  model the magnetic spin fractionalizes into Majorana fermions coupled to a static $\mathbb Z_2$ gauge field.
This allows to map the problem to that of non-interacting Majorana fermions making it an exactly solvable model. In the Kitaev model, the primary excitation of the gauge field is the vison which carries half a flux quantum. Visons are ubiquitous in $\mathbb Z_2$ lattice gauge theories and have been predicted in several systems \cite{visonSenthil,visonoptical,Haovison} but have eluded experimentalists to date. Besides their fundamental importance in predicting signatures of $\mathbb Z_2$ spin liquids, they are much sought after for topological quantum information processing \cite{Kitaev06,toricCode}.

Within the Kitaev model, a vison is an immobile finite-energy excitation, strongly interacting with the gapless Majorana fermions via its flux. The vison should therefore be viewed as a kind of `polaronic' excitation: a $\pi$ flux dressed by a cloud of Majorana fermions. Adding perturbations to the Kitaev model will generically make the gauge field a dynamical degree of freedom with mobile visons.

Remarkably, there are a number of materials which are believed to be approximately described  by the Kitaev model.
The past decade witnessed a surge of experimental efforts to detect fractionalization in such {\em Kitaev materials} \cite{NasuReview,trebstreview,knolle,fieldinduced,rucl3obs,proximate}. Arguably, the most direct evidence so far for an exotic spin liquid phase have
been reports of an approximately half-integer  \cite{kasahara1,Yamashita2020,kasahara2,bruin2021robustness} quantized thermal Hall effect in a magnetic field in $\alpha$-RuCl$_3$
expected to occur in chiral spin liquids coupled to phonons \cite{Balents,RoschQuantumHall}. Recently, very strong oscillations of the longitudinal thermal conductivity have been observed \cite{Ong} and also attributed to fermionic excitations of an exotic spin liquid phase.
Direct experimental signatures of visons, or - more generally - of  emergent dynamical gauge fields, are, however, still missing.
From the theory side, new detection protocols exploiting vison-Majorana interactions in the pure Kitaev limit have been proposed in recent works. This include local probes like STM \cite{egger,knollelocal,udagawa,elio}, interplay of disorder and fractionalization \cite{perkins,knolledisorder}, and spin transport~\cite{Nasu}.

In all real materials the presence of further spin interactions beyond the Kitaev coupling \cite{jackeli,proximate,trebstreview,valenti} is unavoidable.
Such terms, if sufficiently strong, destroy the spin liquid phase, often inducing magnetic ordering. In this case the fractionalized quasiparticles cease to be the most natural description of the model. Several numerical and mean-field studies have investigated the phase diagram of the Kitaev model in the presence of other interactions \cite{YamadaFujimoto,trebstreview,augmented,ciaran,KJG,gohlkeGamma,gohlke2,hykee} and provided useful insights. One interesting feature is, 
for example, that the ferromagnetic Kitaev model turns out to be much more fragile towards perturbations by either an off-diagonal symmetric exchange ($\Gamma$ term) \cite{gohlkeGamma,KJG,hykee} or a magnetic field \cite{ciaran,gohlke2}.
The zero temperature phase transitions triggered by vison-pair (located on two adjacent plaquettes) dynamics have been studied by Zhang and collaborators recently \cite{Batista,zhang}. In a $\mathbb Z_2$ gauge theory, such vison pairs do not carry a net flux. The question whether an isolated vison, which defines due to its fractional flux a singular perturbation for the gapless fermions, is a coherent particle with a well defined mass is a non-trivial question and is largely unexplored. In this paper we provide a controlled calculation of the dynamics of single visons in the limit where perturbations by non-Kitaev terms are weak.

\section{Model}
We consider the isotropic honeycomb Kitaev model \cite{Kitaev06}
in the presence of small perturbations,
\begin{align}
H&=H_K+\Delta H_h+\Delta H_\Gamma+\Delta H_J \\
H_K &=K\sum_{<ij>_\gamma} \sigma^\gamma_i \sigma^\gamma_j.
\end{align}
In the pure Kitaev model, $H_K$, each site on the honeycomb lattice connects to its three neighbors with different components $\gamma=x,y,z$ of the spin.
We mainly focus on two types of perturbations, a magnetic field in the [111]  direction and an off-diagonal symmetric interaction, the so-called $\Gamma$ term
 \begin{align}
\Delta H_h &= -h \sum_{i,\alpha} \frac{(1,1,1)}{\sqrt 3} \cdot \vec{\sigma}_i \\
 \Delta H_\Gamma &= \Gamma \sum_{\langle ij \rangle_\gamma; \alpha,\beta\neq \gamma}\left(\sigma^\alpha_i\sigma^\beta_j+\sigma^\alpha_i\sigma^\beta_j\right).
\end{align}
 Furthermore, we will also comment  on the effects of perturbations arising from an isotropic Heisenberg term $\Delta H_J =J \sum_{<ij> } \vec \sigma_i \cdot \vec \sigma_j$.

The pure Kitaev model can be solved exactly \cite{Kitaev06} by mapping each spin to four Majorana fermion operators $b^x,b^y,b^z$ and $c$ on each lattice site with $\sigma^{\alpha}_i = i b^{\alpha}_i c_i$. The Kitaev Hamiltonian becomes
 \begin{align}
	\label{eqn:maj_ham}
	H = -K \sum_{<ij>_\gamma} i \hat{u}^\gamma_{ij} c_i c_j ,
\end{align}
where  the ``link operators'' $\hat{u}_{ij} = i b^\alpha_i b^\alpha_j$ commute with the Hamiltonian, takes eigenvalues $\pm 1$
and is identified with a $\mathbb Z_2$ gauge field. The honeycomb lattice splits into two sublattices, $A$ and $B$, and in the following we will use a convention
where $i \in A$ and $j \in B$. On each link we define {\em bond fermions} $\chi$  \cite{Baskaran} and in each unit cell {\em matter fermions} 
\begin{align}
	\chi_{\langle ij \rangle_\alpha} = b^\alpha_i+i b^\alpha_j, \qquad f_{i} = c_i+ic_j .
\end{align}
The gauge variable $\hat{u}_{\langle ij \rangle_\alpha} = 2 \chi^\dagger_{\langle ij \rangle_\alpha}\chi_{\langle ij \rangle_\alpha}-1$ now becomes the parity of the bond fermion. 

This spin-Majorana mapping necessarily enlarges the Hilbert space of the original spin model. The {projection} operator $\hat{P}$ is used to project out unphysical states. 
\begin{align}
	\label{eq:D}
	\hat{P} = \prod_k \frac{(1+\hat{D}_k)}{2} ,\qquad	\hat{D}_k = b^x_kb^y_kb^z_kc_k .
\end{align}
From the gauge theoretical perspective, $\hat P$ induces a summation over all $\mathbb Z_2$ gauge transformations.

{\em Visons --} The physical degree of freedom encoded in the $\mathbb Z_2$ gauge field is the flux of each hexagonal plaquette. The plaquette operator $\hat{W}_p=\prod\limits_{\hexagon} \sigma_{i}^\gamma \sigma_j^\gamma=\prod\limits_{\hexagon} u_{ij}$ with eigenvalues $\pm 1$ commutes with $H_K$. In the ground state of $H_K$, $W_p=1$ on all plaquettes describing a flux-free state.
A vison is the gauge excitation with lowest energy obtained by setting one of the $\hat{W}_p=-1$, thus creating a $\pi$ flux. In systems with periodic boundary conditions (PBC) visons can only be created in pairs but
with open boundary conditions (OBC) a single vison is a well defined excitation \cite{Kitaev06} with a finite energy cost $E^v_0 \approx 0.1535 |K|$. 

Within the gauge theory description, one can describe a vison by a string of flipped link variables $u_{ij}=-1$. This string extends to the boundary (OBC) or connects a pair of visons (PBC). To handle this unphysical gauge string while calculating gauge invariant quantities, we find it useful to project the wave functions back to the physical Hilbert space.
\begin{align}
	\label{eq:spin_wf}
	\ket{\Phi(\boldsymbol{R})} =  \hat{P}\ket{\mathcal{G}(\boldsymbol{R})}\ket{M(\mathcal{G})} .
\end{align}
Here $\vec R$ denotes the position of the vison, $\ket{\mathcal{G}(\boldsymbol{R})}$ is the wavefunction describing the gauge sector (i.e., the bond fermions) while $\ket{M(\mathcal{G})}$ is the many-body wavefunction of the Majorana fermions in a fixed gauge $\mathcal G$. Importantly, $\hat P$ projects the wavefunction onto the physical Hilbert space.

To avoid numerical problems related to dangling bonds and spurious boundary modes, we do all of our calculations with periodic boundary conditions, placing two visons at maximal separation. Using exact diagonalization, we typically consider systems with linear dimensions up to 80 corresponding to 12.800 sites.

\section{FM Kitaev}
\subsection{\new{Linear Perturbation theory}}
 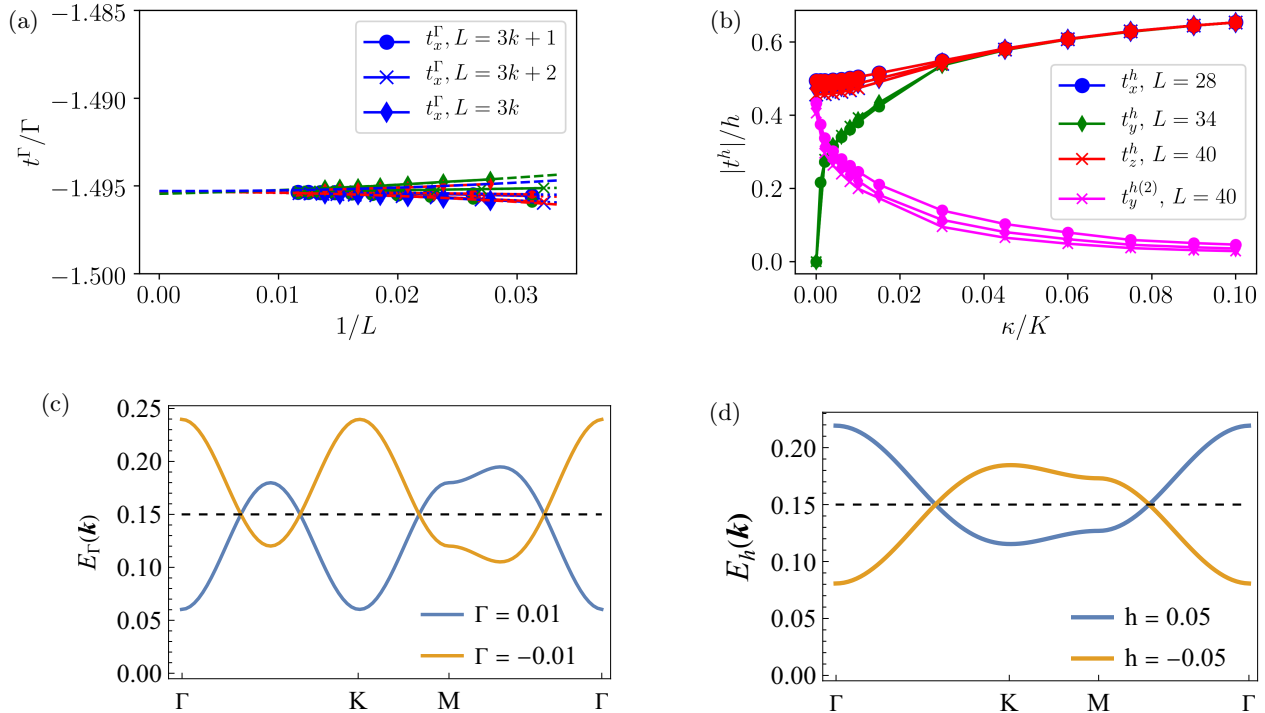
\begin{figure*}
	\centering
	\begin{tikzpicture}
	\node[anchor=center] (image) at (0,0) {\includegraphics[width=0.45\textwidth]{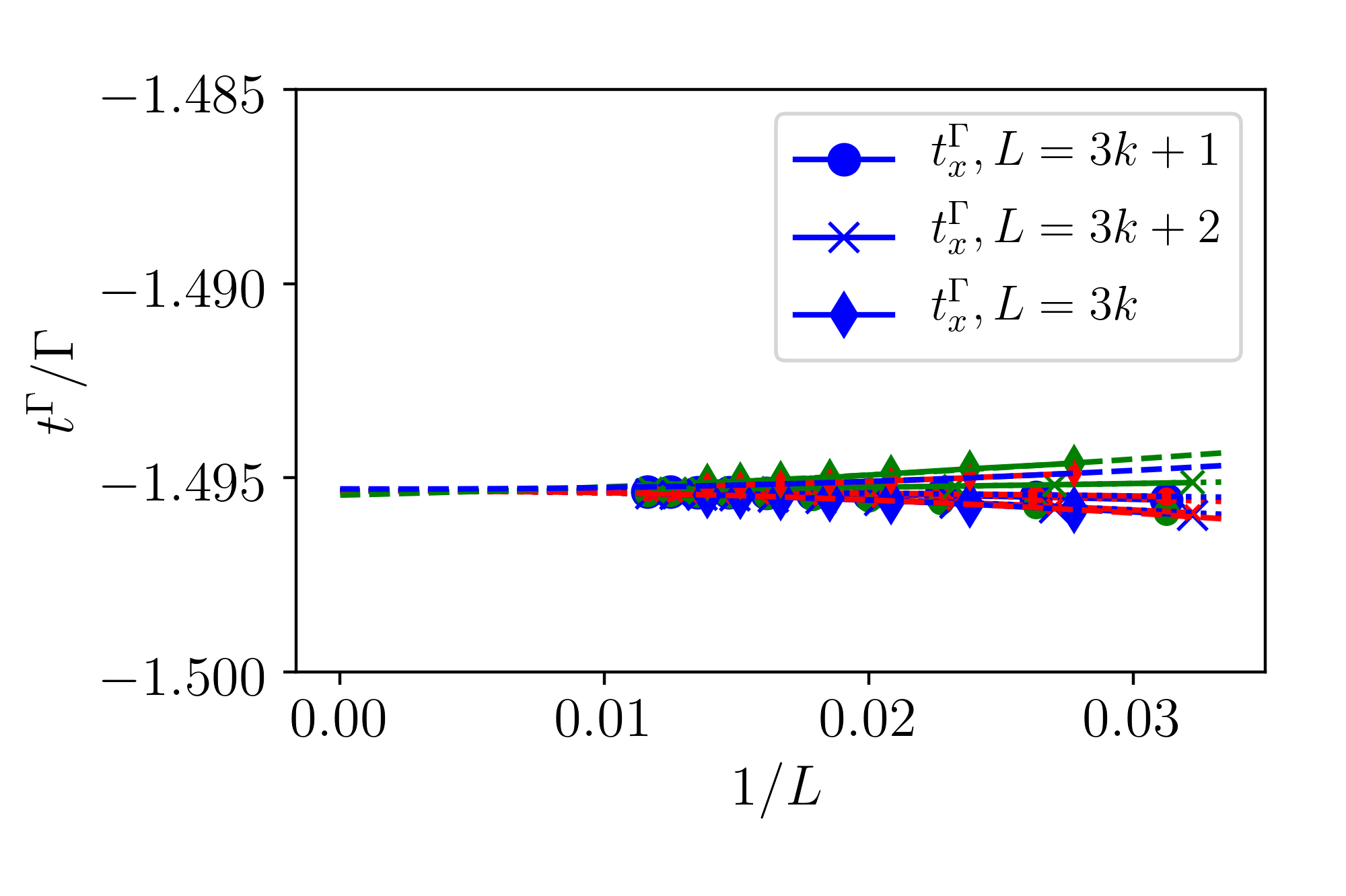}};
	\node[anchor=center] (text) at (-3.8,2) {\small{(a)}};
	\end{tikzpicture}
	\qquad
	\begin{tikzpicture}
	\node[anchor=center] (image) at (0,0) {\includegraphics[width=0.45\textwidth]{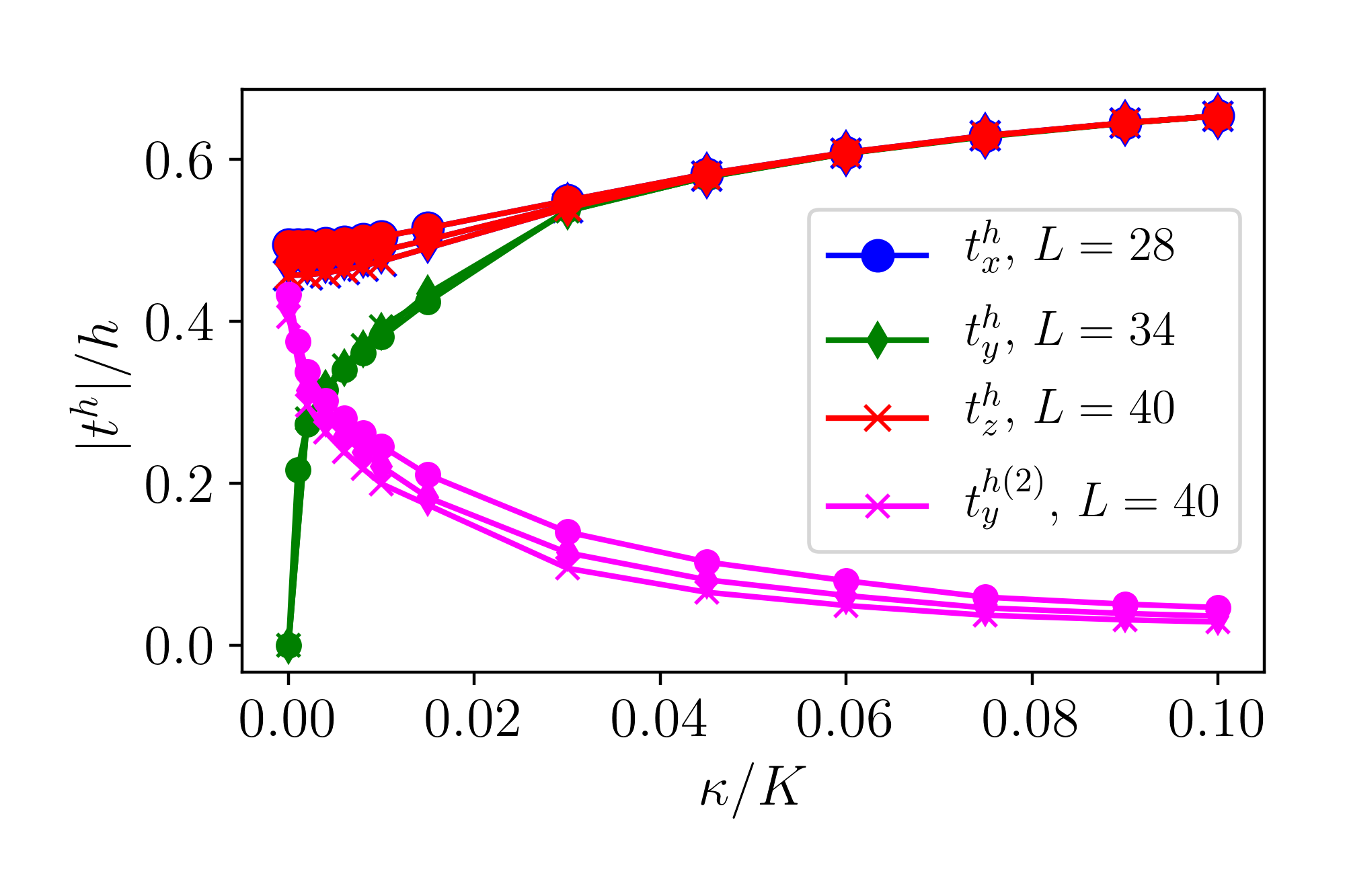}};
	\node[anchor=center] (text) at (-3.5,2) {\small{(b)}};
	\end{tikzpicture}

	\begin{tikzpicture}
	\node[anchor=center] (image) at (0,0) {\includegraphics[width=0.4\textwidth]{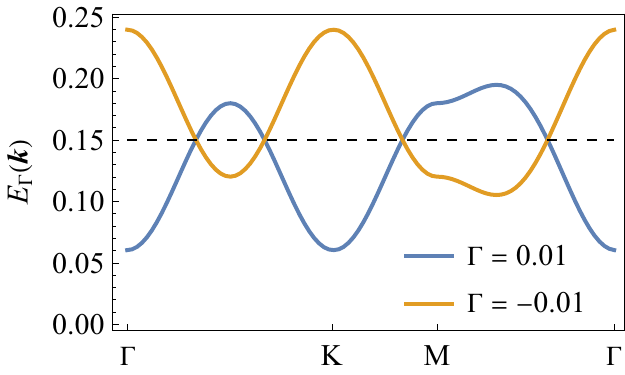}};
	\node[anchor=center] (text) at (-3.8,2) {\small{(c)}};
	\end{tikzpicture}
	\qquad \quad
	\begin{tikzpicture}
	\node[anchor=center] (image) at (0,0) {\includegraphics[width=0.4\textwidth]{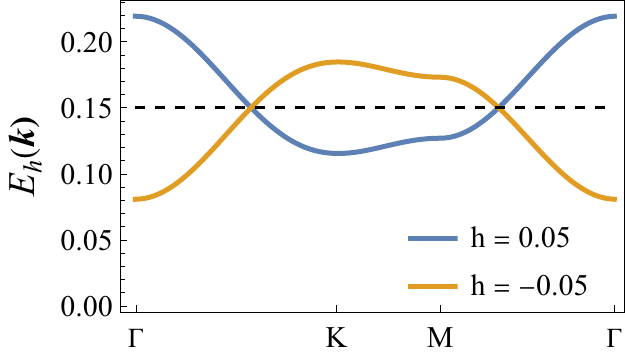}};
	\node[anchor=center] (text) at (-3.5,2) {\small{(d)}};
	\end{tikzpicture}
	\caption{\small (a.) Vison hopping amplitudes for $K=-1$ as function of inverse system size, $L=3 k + n$, $k \in \mathbb N$, $n=0,1,2$ for a perturbation by a small $\Gamma$ term (next-nearest neighbor hopping, $\kappa=0$) (b.) Vison hopping amplitude (magnitude) induced by a small magnetic field $h$ for $K=-1$ as a function of Majorana mass gap $\kappa$. The magenta plot shows the hopping from a ground-state to an excited state of a nearest neighbour site. , Color code: green - $t^\zeta_y$, red - $t^\zeta_z$ where $\zeta = h, \Gamma$. In panel (c) and (d) the corresponding vison dispersions are shown.\label{fig:sub1}}
\end{figure*}

 We now turn to the case with small perturbations $\Delta H = \Delta H_\Gamma,  \Delta H_h$. These terms obviously break the exact solubility of the pure Kitaev model as the plaquette operators are no more conserved. Thus the gauge field becomes a dynamical degree of freedom, visons are created and destroyed by quantum and thermal fluctuations and they become mobile. Importantly, the vison number remains conserved modulo $2$ and thus a single vison cannot decay but remains a stable quasiparticle. To linear order in the perturbations, the hopping rate of the vison can be computed from
 \begin{align}
	t_{ab} =  \bra{\Phi_0(\boldsymbol{R}_a)}\Delta H \ket{\Phi_0(\boldsymbol{R}_b)}\label{eq:matrixHoppig} .
\end{align}
The second vison in our system is kept at a fixed position, while computing the hopping from vison position $\vec R_b$ to $\vec R_a$.
The computation of this harmless-looking overlap, discussed in App.~\ref{App:me},  turns out to be non-trivial for three reasons. First, it is important to 
use the projection operator $\hat P$ in Eq.~\eqref{eq:D} to be able to match different gauges. Second, one has to calculate fermionic matrix elements involving the overlap of two different many-particle Majorana states and corresponding Bogoliubov vacua which can be done using methods developed by Robledo~\cite{Robledo,robledo2}. Third, some (but not all) of the matrix elements have strong finite size effects probably related to the presence of a gapless spectrum and quasi-localized states induced by the vison \cite{siteDilution,vacancyDOS}.
For the ferromagnetic Kitaev model, $K<0$, the $\Gamma$ term induces a next-nearest neighbor hopping $t_\Gamma$ of the vison (on the dual triangular lattice formed by the plaquettes). Fig.~\ref{fig:sub1}a shows that finite size effects are almost absent and
we obtain
\begin{align}
t_\Gamma \approx -1.495 \, \Gamma .
\end{align}
In Fig.~\ref{fig:sub1}c, the resulting band structure is shown.  For $\Gamma>0$ there are 6 minima located on the lines connecting the $\Gamma$ and $M$ points. For $\Gamma<0$, the minima of the dispersion are located the $\Gamma$, $K$ and $K'$ points. That the energy at the $\Gamma$ point is exactly the same as at the $K$ and $K'$ points is an artifact of our leading-order approximation 
which includes only next-nearest neighbor hopping.

\new{An external magnetic field $h$ in the $(111)$ direction has two effects: to linear order in $h$ it induces a hopping of the vison, to cubic order a gap of size $2 \kappa \propto \frac{h^3}{K^2}$ is opened \cite{Kitaev06} in the Majorana spectrum (here we assume $\Gamma=0$ \cite{Balents}). While this scaling suggests that one can simply ignore the effects of $\kappa$ to lowest order perturbation theory,
the presence of a Majorana zero mode attached to the vison for $\kappa\neq 0$ (or a quasi-bound state for $\kappa=0$) makes the analysis more subtle and induces strong finite size effects.

In Fig. \ref{fig:sub1}.b we show the amplitude of magnetic field induced vison hopping for three different directions (across $x$, $y$ and $z$ bonds) as function of Majorana gap $\kappa$. Besides the ground-state to ground-state hopping rates, it turns out that in the small $\kappa$ limit one has also to include the hopping to an excited state  (with energy $E_V+2\kappa$ where $E_V$ is the ground state energy of the vison) for certain directions of hopping.

The results depend on the ratio of two length scales, the distance between the two visons $d_V$ and the extend of the Majorana bound state attached to the vison, $\xi_m \sim v_m/\kappa$. For $d_V \gg \xi_m$ (corresponding to $\kappa>0.03\, |K|$ in Fig. \ref{fig:sub1}) one can ignore the hopping to the excited state and one obtains a finite, directionally independent hopping rate of the vision with almost no finite size effects and only a weak dependence on $\kappa$. For example, for $\kappa=0.05\,|K|$ we find 
\begin{align}
|t_h| \approx 0.6\,h
\end{align}

In the opposite limit, $d_V \lesssim \xi_m$ (small $\kappa$ limit in Fig. \ref{fig:sub1}b), in contrast, we obtain very large finite size effects and the hopping rates across the $y$ bonds 
become different from those across the $x$ and $z$ bonds of the Kitaev lattice. This is a consequence of the presence of the second vison which explicitly breaks the rotational symmetries. Furthermore, in the small $\kappa$ limit one cannot ignore the hopping $t_y^{h(2)}$ to excited states (magenta lines in Fig.~\ref{fig:sub1}b) across the $y$ bond which becomes much larger than the groundstate-to-groundstate hopping $t_y^{h}$ (green line) for $\kappa\to 0$. The case $\kappa=0$ is special and highly singular ($t_y^h=0$ and $t_y^{h(2)}\approx t_x^h \approx t_z^h$). As detailed in  Appendix.~\ref{App:me}, in this case the relative fermionic parity of the  states appearing in Eq.~\eqref{eq:matrixHoppig} depends in a non-trivial way on the position of the second vison. Thus certain hopping processes are only allowed if an extra matter Majorana mode is occupied.

This analysis shows that the very notion of a single and independent vison excitation is {\em not} well defined in the limit when the vison-vison distance $d_V$ is smaller than $\xi_m$. In this case one cannot formulate a theory of a single vison because the (quasi-) bound Majorana state attached to one vison interacts with neighboring visons.

In contrast, for $d_V \gg \xi_m$, one can treat a single vison as a well-defined independent particle. Remarkably, our calculation shows that the situation is also different for the $\Gamma$ perturbation: in this case the single-vison hopping is with high precision independent of the presence of the second vison. Thus it is possible to formulate a theory of single visons also in this case even for a gapless Majorana spectrum (see also Sec. \ref{mobility} below).

In Fig.~\ref{fig:sub1}d we show the vison dispersion for $d_V \gg \xi_m$ for a finite gap $m$ in the Majorana spectrum. In the ferromagnetic Kitaev model discussed here (and in contrast to the antiferromagnetic case discussed in Sec.~\ref{sec:afm}), the vison hopping rates can be chosen to be real. This means none of the vison lattice plaquettes enclose a non-zero flux and the vison bands carry no Chern number.
}

\subsection{\new{Vison Mobility}} \label{mobility}
\begin{figure}[b]
	\centering
	\includegraphics[width=0.43\textwidth]{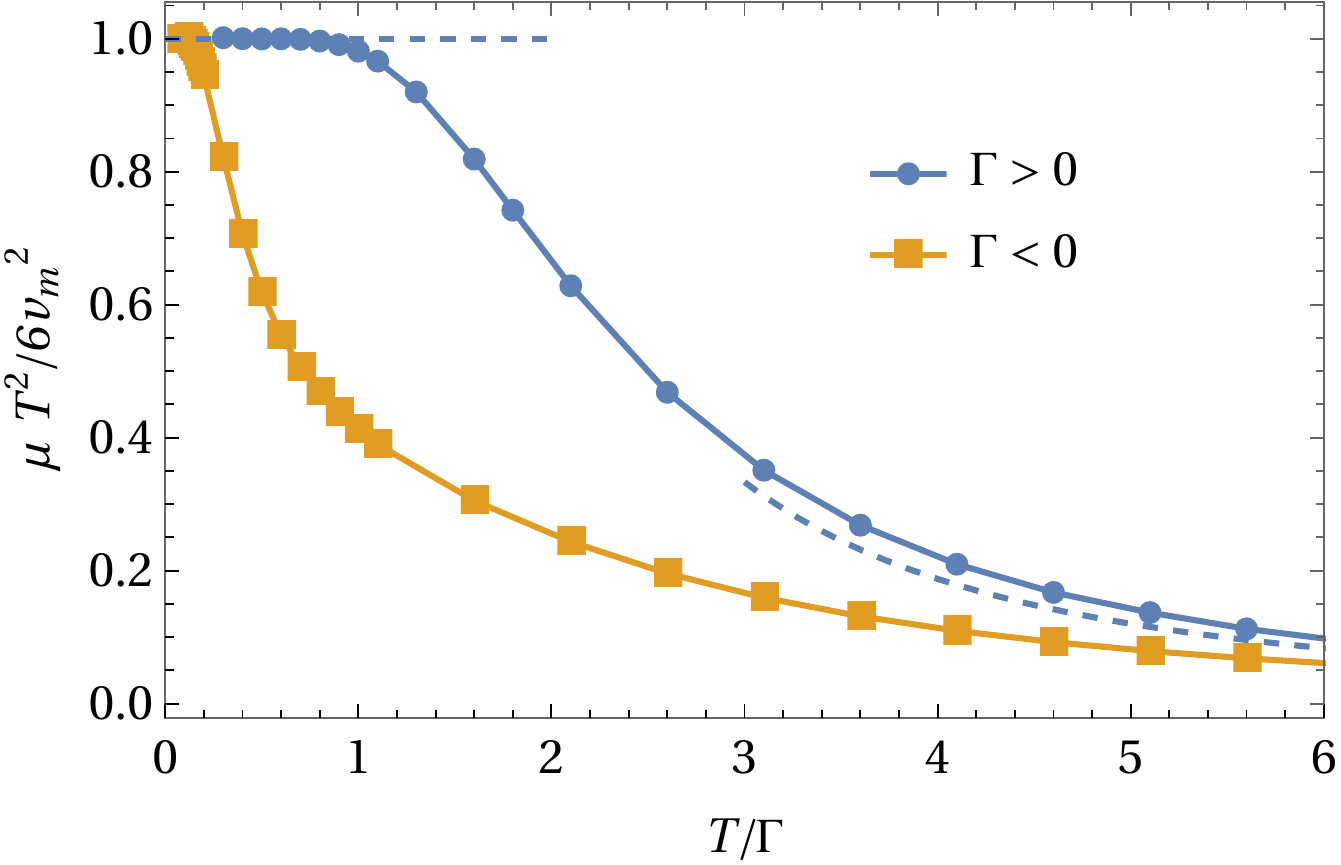}
	\caption{Vison mobility, $\mu/(6 v_m^2/T^2)$, in the ferromagnetic Kitaev model perturbed by a $\Gamma$ term. $\mu$ is normalized to its low-$T$ asymptotics and plotted as function of $T/|\Gamma|$ both for $\Gamma>0$ and $\Gamma<0$. Deviations from the universal low-$T$ mobility are more pronounced for $\Gamma<0$ at low $T$ due to flat regions in the band structure close to the band minimum, see Fig.~\ref{fig:sub1}. The dashed lines indicate the low-$T$ and high-$T$ asymptotics, see Eq.~\eqref{eq:muT}. \label{fig:mobility}}
\end{figure}
So far we have shown that a dressed vison obtains a finite hopping amplitude linear in $h$ and $\Gamma$ at zero temperature. At finite temperatures, thermally excited gapless Majoranas will scatter from the vison, leading to friction and a finite mobility of the vison. The mobility $\mu$ describes the finite velocity $v$ obtained by a vison in the presence of external forces, $\langle v \rangle = \mu F$. Via the Einstein relation $D=\mu k_B T$ the mobility is directly related to the diffusion constant of the vison which characterizes its dynamics. Note that calculation of the mobility of a vison is qualitatively different from the problem of the mobility of a vortex in a d-wave superconductor where extra complications arise due to the presence of Goldstone modes and the external magnetic field \cite{volovik,vinokur,sachdevVortex}. Here, we consider the effect of the $\Gamma$ perturbation for $K<0$ and comment on the applicability of our results for other situations below.

We consider the limit, where the temperature $T$ is smaller than the vison gap (so that the density of visons is small).
 In this regime, we can describe the Majorana modes by a Dirac equation with velocity $v_m$. The scattering cross section of 2D Dirac electrons from a $\pi$ flux is well known \cite{durst,AB} (see also App.~\ref{App:B}) and given by
$\frac{d \sigma}{d \theta}=\frac{1}{2 \pi k \sin^2(\theta/2)}$.
Furthermore, we can use that the momentum transfer $\Delta p \sim T/v_m$ during a scattering process is small compared to the typical vison momentum $\sim \sqrt{T/W_v}/a$, where $W_v=9 |t_\Gamma|$ is the vison bandwidth and $a$ the lattice constant. As shown in App.~\ref{App:mobility}, this allows to rewrite  \cite{howell2004} the singular Boltzmann scattering kernel into a non-singular drift-diffusion equation in momentum space describing Brownian motion. 
	\begin{align}
	\partial_t f_\vec{p}+\boldsymbol v^v_\vec{p}\cdot \boldsymbol F \frac{d f^0}{d E^v_\vec{p}} \approx D_p \left( \nabla^2_{\vec p} f_{\vec p}+\frac{1}{T} \, \vec \nabla_\vec{p} \left( \vec v^v_\vec{p}f_{\vec p}\right)\right)\label{eq:driftdiffusionfmain},
\end{align}
where $f_\vec{p}$ is the vison distribution function, $v^v_\vec{p}=d E^v_\vec{p}/d \vec p$ the vison velocity and $D_p=6 T^3/v_m^2$ is the diffusion constant in momentum space, see  App.~\ref{App:mobility}.
The asymptotic behaviour of the mobility can then be calculated analytically 
\begin{align}
\mu(T)= \frac{D(T)}{T}=\left\{ \begin{array}{ll} \frac{18 t_\Gamma^2 v_m^2}{T^4} & \text{for } K \gg T\gg W_v \\[2mm] \frac{6 v_m^2}{T^{2}} & 
\text{for } T\ll W_v\end{array}\right. .\label{eq:muT}
\end{align}
Remarkably, the low-temperature mobility $\mu(T)$ and therefore also the vison diffusion constant $D(T)$ are fully universal and completely independent of the vison dispersion, which follows from the
scale invariance of the problem and the universal scattering cross section.
Similar results (with different prefactors) exist for the problem of a vortex in a d-wave superconductor \cite{sachdevVortex}. In Fig.~\ref{fig:mobility} we show the mobility as function of $T$ for different values of $\Gamma$.

Above we only considered the effect of a small $\Gamma$ term for $K<0$. However, the same universal low-$T$ mobility and the same $T$ dependence at larger $T$ is expected for arbitrary vison bands as long as (i) the vison bandwidth is small compared to the Majorana bandwidth, (ii) their dispersion is quadratic at the bottom of the band and (iii) the Majorana dispersion can be described by a Dirac equation. Thus, in the case of magnetic field, the formula for the mobility is only valid for temperatures large compared to the field-induced gap in the Majorana spectrum.

\section{AFM Kitaev }\label{sec:afm}
\subsection{\new{First order Perturbation theory}}
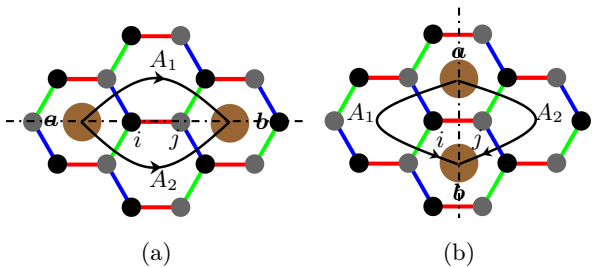
\begin{figure}[b]
 	\begin{tikzpicture}
 	\node[anchor=center] (image) at (0,0) {\includegraphics[width=0.2\textwidth]{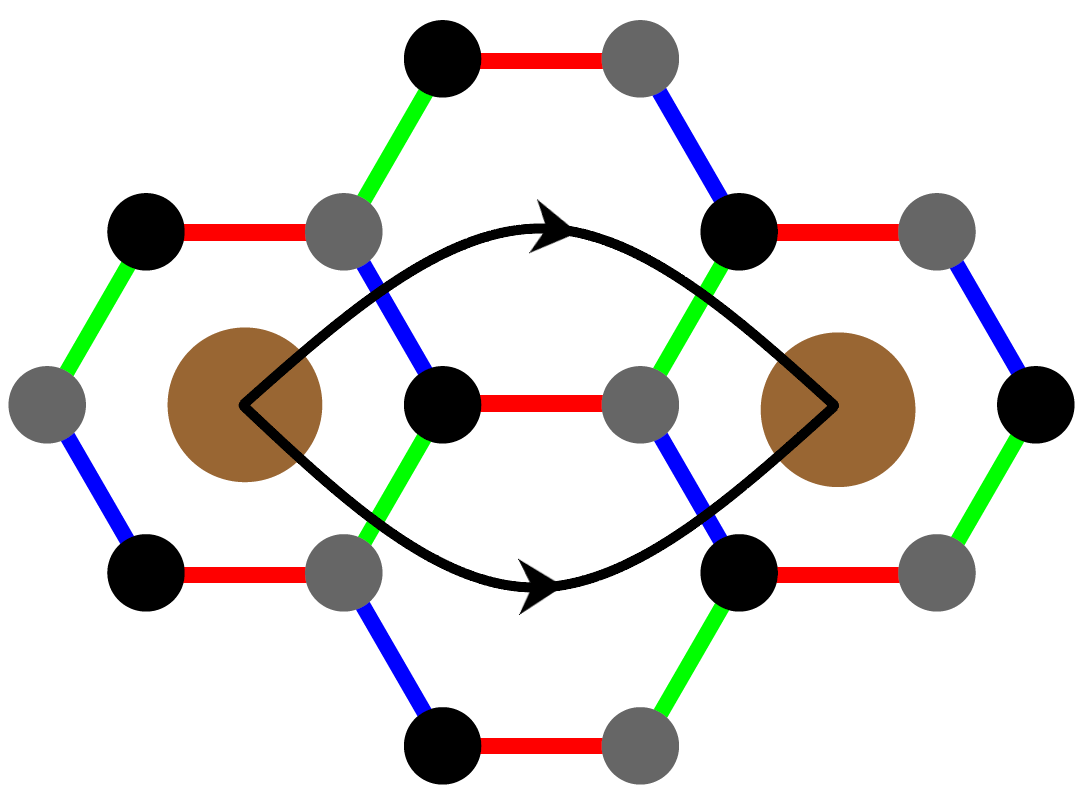}};
 	\node[anchor=center] (text) at (-.25,-.25) {\small $i$};
 	\node[anchor=center] (text) at (0.25,-0.25) {\small $j$};
 	\node[anchor=center] (text) at (-1.4,0) {\small $\boldsymbol{a}$};
 	\node[anchor=center] (text) at (1.4,0) {\small $\boldsymbol{b}$};
 	\node[anchor=center] (text) at (0.1,0.8) {\small $A_1$};
 	\node[anchor=center] (text) at (.1,-0.8) {\small $A_2$};
	\draw [thick,dash dot] (-2,-0.01) -- (2,-0.01);
 	\node [below=1.5cm, align=flush center,text width=2cm] at (image)
 	{
 		(a)
 	};
 	\end{tikzpicture}
 	\begin{tikzpicture}
 	\node[anchor=center] (text) at (0,0) {\includegraphics[width=0.2\textwidth]{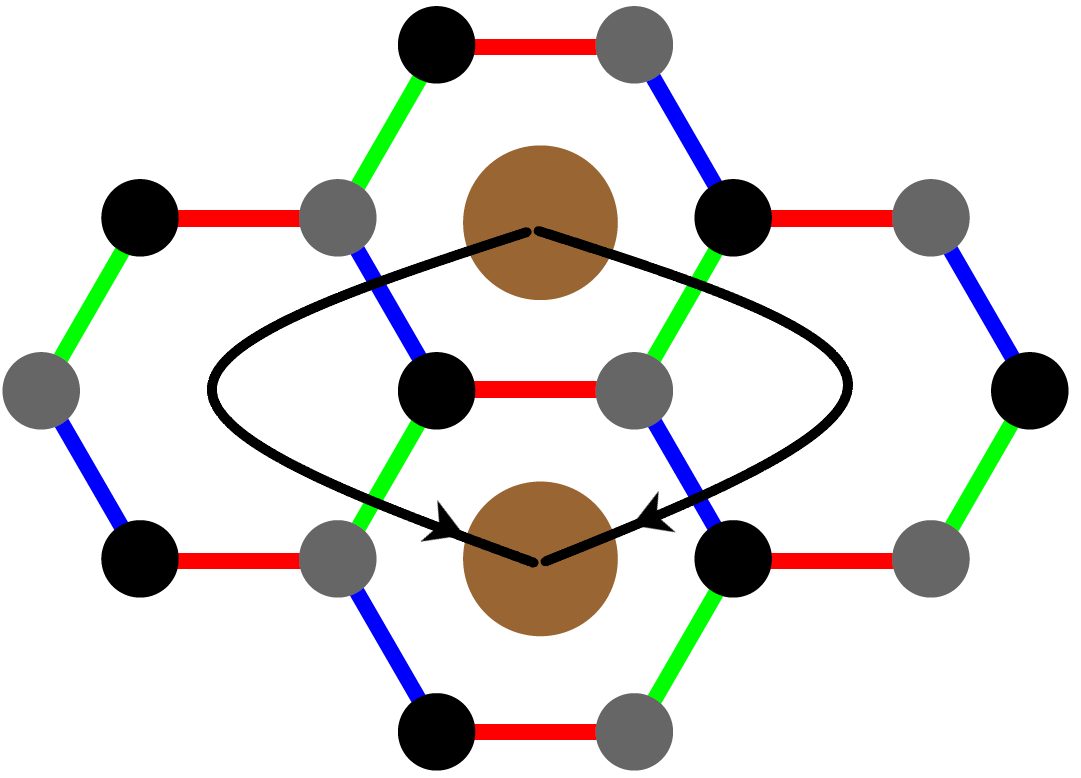}};
 	\node[anchor=center] (text) at (-.25,-.25) {\small $i$};
 	\node[anchor=center] (text) at (0.25,-0.25) {\small $j$};
 	\node[anchor=center] (text) at (-1.3,0.1) {\small $A_1$};
 	\node[anchor=center] (text) at (1.2,0.1) {\small $A_2$};
 	\node[anchor=center] (text) at (0,0.9) {\small $\boldsymbol{a}$};
 	\node[anchor=center] (text) at (0,-0.95) {\small $\boldsymbol{b}$};
 	\draw [thick,dash dot] (0,1.5) -- (0,-1.3);
 	\node [below=1.5cm, align=flush center,text width=2cm] at (image)
 	{
 		(b)
 	};
 	\end{tikzpicture}
 	\caption{Vison hopping processes induced by (a) $\Delta H_\Gamma$ (b) $\Delta H_h$. The brown disks represent the visons (positions $\boldsymbol{R}_a$ and $\boldsymbol{R}_b$) and the black curves show different trajectories that interfere constructively (destructively) for FM (AFM) Kitaev interaction.\label{fig:interference}}
 	\vspace{-2em}
 \end{figure}

When evaluating the vison hopping rate, Eq.~\eqref{eq:matrixHoppig}, for a antiferromagnetic Kitaev coupling, $K>0$, 
we obtain the remarkable result that it vanishes exactly for both $h$ and $\Gamma$ perturbations \new{in the limit of vanishing Majorana mass gap $\kappa$}. To understand the origin of this effect, it is useful to realize that a single vison hopping process arises from the interference of two contributions, $t_{ab}=A_1+A_2$ due to two different terms in the Hamiltonian $\Delta H_1$ and $\Delta H_2$.
For example, for the $z$-link shown in Fig.~\ref{fig:interference}, $\Delta H_1=\Gamma \sigma^x_i\sigma^y_j$ (or $\Delta H_1=\frac{h}{\sqrt{3}} \sigma^z_i$) while $\Delta H_2=\Gamma \sigma^y_i\sigma^x_j$ (or $\Delta H_2= \frac{h}{\sqrt{3}} \sigma^z_j$).
Importantly, these two terms are related by a  reflection symmetry (dashed lines in Fig.~\ref{fig:interference}), which ensures that $A_1=\pm A_2$. To fix the sign, we observe that
	$\langle \Delta H_1 \Delta H_2 \rangle =\langle \sigma^z_i\sigma^z_j\rangle$
  is negative in the AFM Kitaev model while positive in the FM Kitaev model.
This strongly suggests that $A_1=-A_2$ in the AFM phase  as we confirmed numerically by direct evaluation of Eq.~\eqref{eq:matrixHoppig}: a destructive interference eliminates the leading vison hopping process.
 \begin{align}
 \lim_{\kappa \to 0}t_h=\lim_{\kappa \to 0}t_\Gamma=0.\label{zeroHopp}
 \end{align}
 This effect is reminiscent of the `Aharonov-Bohm caging' describing the localization by destructive interference which 
 is often induced in models with $\pi$-fluxes and nearest-neighbor hopping only \cite{ABcage,rizziABcage}. 
 Note that longer-range hopping arising to quadratic orders in $h$ or $\Gamma$ may still possible in our system.

 \begin{figure*}
	\centering	
	\begin{tikzpicture}
	\node[anchor=center] (image) at (0,0) {\includegraphics[width=\textwidth]{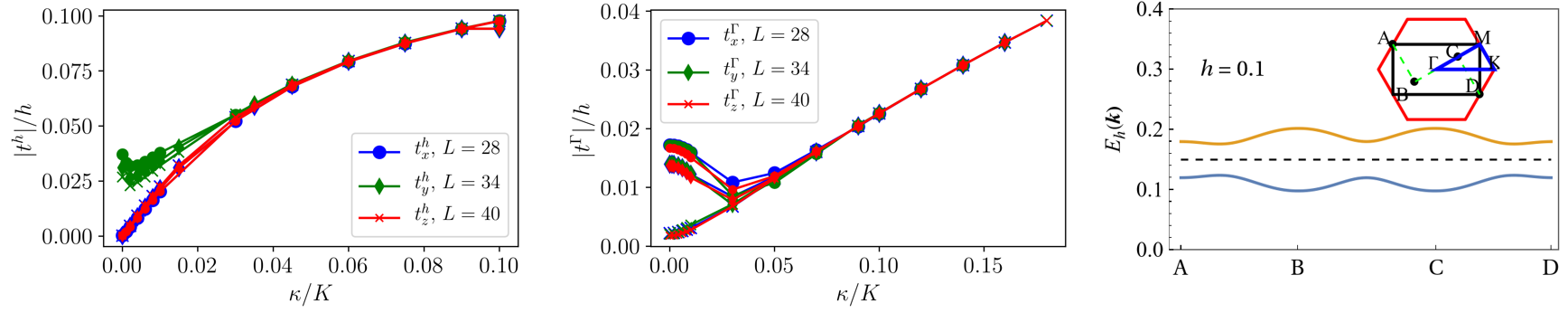}};
	\node[anchor=center] (text) at (-8.5,2) {\small{(a)}};
	\node[anchor=center] (text) at (-2,2) {\small{(b)}};
	\node[anchor=center] (text) at (4,2) {\small{(c)}};
	\end{tikzpicture}
	%

	\caption{\small  \textbf{AFM Kitaev model}: Vison hopping amplitudes for $K=1$ as function of Majorana gap $\kappa$ for a perturbation by a small magnetic field $h$ for different system sizes (panel (a),  Color code: blue-$t^h_x$, green - $t^h_y$, red - $t^h_z$. The phases of the hoppings are such that every triangular plaquette of the vison lattice carries a flux of $-\pi/2$ for $h>0$. In panel (c) the resulting vison bands with Chern number $\pm 1$ are plotted.\label{fig:afm}}
\end{figure*}
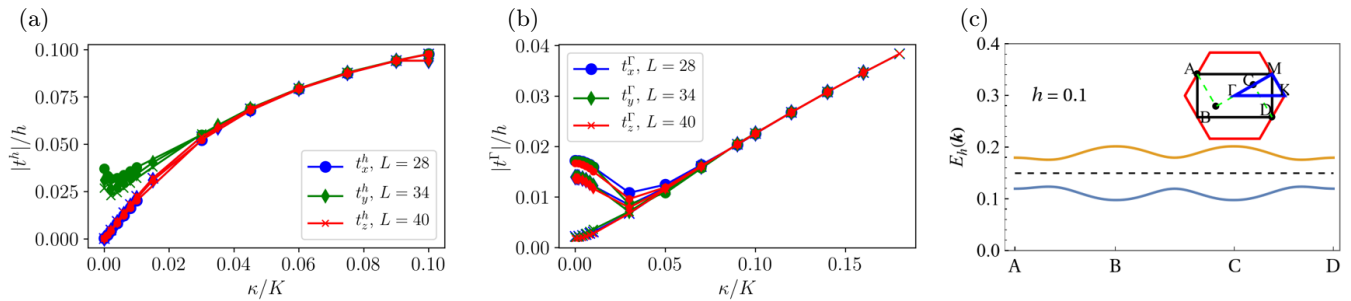
\new{In the presence of an external $(111)$ field, however, it is important\cite{Intiafm} to take into account that $h$ also opens a gap $2\, \kappa$ in the Majorana sector with $\kappa = h^3/K^2$ for $\Gamma=0$ \cite{Kitaev06}. Note that $\kappa \propto h$ when  both Heisenberg and $\Gamma$ perturbations are present \cite{Balents}. Importantly, $\kappa$  breaks the mirror symmetries which led to the destructive interference of vison hopping paths discussed above. Thus, in the presence of $\kappa$, both the field-induced hopping rate $t_h$ and the $\Gamma$ induced rate $t_\Gamma$ become finite. In Fig. \ref{fig:afm}, we plot these hopping amplitudes as function of mass $\kappa$ for different vison separations ($d_v=L/2$). 
In Fig. \ref{fig:afm}.a we can see similar finite size effect as in the FM model (Fig.\ref{fig:sub1}) where the second vison breaks the rotation symmetry in the small mass limit. For $d_V \gg \xi$, finite size effects are, however, absent. For $\kappa=0.05$, for example, we find 
\begin{align}
|t_h| \approx  0.07 \, h\label{eq:th_AFM} .
\end{align}
Our numerical data is roughly consistent with \begin{align}|t_h|\approx 0.32 \, h \sqrt{| \kappa/K|} \label{eq:th_afm2} \end{align}
 in the regime $d_V \gtrsim \xi$ but a reliable extraction of the powerlaw in  $\kappa$ is not possible from our data.

We also determine the phase acquired by the vison around a triangular plaquette, by calculating $\arg\left[\langle \vec R_1|\Delta H_h|\vec R_3 \rangle 
\langle \vec R_3|\Delta H_h|\vec R_2 \rangle 
\langle \vec R_2|\Delta H_h|\vec R_1 \rangle\right]=-\text{sign}(h) \frac{\pi}{2} $ for three vison sites ordered anticlockwise around a honeycomb site.
Thus each triangular vison plaquette (i.e, each site of the original honeycomb lattice) carries a flux of $-\pi/2$ for $h>0$ ($\frac{\pi}{2}$ for $h<0$). Ref.\cite{inti} found a flux of $\pi$ for a vison transported around a unit cell of the honeycomb lattice, consistent with our calculation. 
This leads to a doubling of the unit cell (containing two triangular plaquettes each) and results in two vison bands in a reduced Brillouin zone (see Fig.~\ref{fig:afm}.c), with non-trivial topology characterised by Chern numbers $\pm 1$. This leads to a remarkable prediction that not only the matter Majornanas but mobile visons can also contribute to thermal Hall effect discussed below.}

\new{If an external magnetic field induces a finite mass term $\kappa$, also the interference effect which suppressed $\Gamma$-induced hopping
is affected. In Fig.~\ref{fig:afm}.b we show that the $\Gamma$ induced hopping is linear in $\kappa$ in this case,
\begin{align}
|t_\Gamma|\approx 0.2\, \Gamma \frac{|\kappa|}{K} \label{eq:th_afm} 
\end{align}
Within our perturbative approach it is unlikely that this term dominates: for small $h$ and thus small $\kappa$, higher-order terms in $\Gamma$, $t_\Gamma \sim \Gamma^2$ will dominate, while for larger $h$, one reaches the regime where $|t_h| > |t_\Gamma|$.
}

 \subsection{\new{Majorana-assisted hopping}}
The perfect destructive interference, which prohibits vison motion linear in $\Delta H$ in the AFM case, is disturbed when the vison scatters from thermally excited Majorana fermions. Thus at $T>0$ there will be
a Majorana-assisted {\em incoherent} hopping process with rate $W$.
As $\Delta H$ is small, we can use Fermi's golden-rule to compute the hopping rate $W$ for a vison moving from
site $\vec R_a$ to $\vec R_b$. The fact that the presence of the vison strongly disturbs the Majorana fermions makes this a non-standard calculation. We can use, however, that
for $T \ll K$ the Majorana density is low and the calculation can be done in a continuum model 
describing the vison by a point-like $\pi$ flux, see App.~\ref{App:assisted} for details.
\begin{align}
W =2 \pi \sum_{k,k',l,l'} |\langle k',l',R_b| \Delta H |k,l,R_a\rangle|^2 n(\epsilon_k) \delta(\epsilon_{k}-\epsilon_{k'}) \label{eq:goldenRule}
\end{align}
Here $l,l'$ are the angular momentum quantum numbers of the scattering wave functions, $n(\epsilon_k)$ is the Fermi function and
$\epsilon_k=v_m k$ the dispersion of low-energy Majoranas.
The hopping rate $W$ induces a random walk on the vison lattice, from which the diffusion constant $D$ and thus (via Einstein's relation) the mobility can be obtained.
$W$ and thus $D$ are linear in $T$, see App.~\ref{App:mobility}, therefore we obtain a $T$-independent mobility
\begin{align}
\mu(T) = \frac{D(T)}{T} \sim \left\{\begin{array}{ll}  \frac{\Gamma^2  a^4}{v_m^2} &  \text{for } K \gg T \gg \sqrt{\Gamma K}\\[2mm]
   \frac{h^2 a^4}{v_m^2} &  \text{for } K \gg T \gg \sqrt{h K} \end{array} \right. \label{eq:mu_assist}
\end{align} 
for perturbations by $\Gamma$ and $h$, respectively. The formula is valid only for rather high temperatures, because
at lower $T$ coherent second-order (longer-range) hopping processes set in giving rise to a bandwidth of order $W_v^{(2)}\sim\Gamma^2/K, h^2/K$. In the low-temperature regime, one can simply replace $W_v$ and $t_\Gamma$ by $W_v^{(2)}$ in Eq.~\eqref{eq:muT}
to obtain an estimate for the mobility.

The $T$-independent mobility of Eq.~\eqref{eq:mu_assist} is reminiscent of ohmic friction, but its physical origin (assisted hopping) is very different compared to, e.g., Landau damping.
\section{Heisenberg interaction}
Finally, we briefly discuss the effects of a small perturbation by a Heisenberg term, 
$\Delta H_J= J \sum_{ij} \vec \sigma_i \vec \sigma_j$. Applying $\Delta H_J$ to a single vison creates a state with three or five visons. Thus there is no vison hopping linear in $J$. While we have not performed a complete calculation to 
order $J^2$, we argue in App.~\ref{App:heisenberg} that single-vison hopping processes at order $J^2$ cancel by an interference effect very similar to the one discussed above for $K>0$. An important difference is, however, that this destructive interference occurs for both signs of $K$. This suggests that coherent
 vison hopping induced by $J$ may occur only to 
order $J^4$. In contrast, a bound vison pair ($b$ fermions) can hop already to linear order in $J$ as recently shown by Zhang {\it et al.} \cite{Batista}.
For single-vison hopping, however, we expect that $\Gamma$ is much more important than $J$.

\section{Experimental signatures of mobile visons}\label{halleffect}
\new{The motion of visons is expected to affect practically all physical properties and observables of Kitaev materials.
In most spectral probes, however, it will simply lead to an extra broadening of spectra.
On a more qualitative level, vison motion breaks the integrability of the system and allows it to thermalize.
Consider, for example, the transition from a state with a finite density of single visons (e.g., after heating the system with a laser) to a state with zero (or much lower) vison density. Without vison motion such a system cannot equilibrate and thus the vison motion is expected to be the bottleneck for equilibriation. For vison distances large compared to the vison-Majorana scattering length, the motion of visons is diffusive and thus the time-scale $\tau_{VV}$ for two visons to meet is set by $\tau_{VV} \sim \Delta_V^2/D =1/(D n_V)$, where $D$ is the vison diffusion constant, $\Delta_V$ is a typical vison-vison distance and $n_V$ is the vison density.
Thus, the vison-vison annihilation is expected to obey the equation
\begin{align}
\partial_t n_V = - \alpha D n_V^2
\end{align}
where $\alpha$ is the (dimensionless) probability that two  visons, which meet, annihilate each other. 
We have checked the validity of this phenomenological equation for a simple two-dimensional random-walk toy model of diffusing particles which annihilate when they meet. This equation is solved by $n_V(t)=\frac{n_0}{1+\alpha D n_0 t}$. Thus for time scales large compared to the initial vison-vison annihilation time, one obtains the remarkably simple and universal result
\begin{align}
n_V(t) \approx \frac{1}{\alpha D t} \quad \text{for } t \gg \frac{1}{\alpha D\, n_V(t=0)}.
\end{align}
We thus expect that a characteristic $1/t$ tail will show up in pump-probe experiments at low temperatures, with a prefactor governed by the diffusion constants of Eq.~\eqref{eq:muT} with $D=6 \hbar v_m^2/k_B T$ in the low-$T$ regime. Note that $1/t$ long-time tails (typically with very small prefactors) also exist in two-dimensional systems with conservation laws \cite{RoschMitra} but here the vison density is not conserved (and energy can be transported from layer to layer by phonons in 3d experimental systems like  $\alpha$-RuCl$_3$).

A striking result is the emergence of vison bands with finite Chern numbers in the antiferromagnetic Kitaev model. This will lead to an extra contribution to the thermal Hall effect (THE). Note that any vison contribution to the THE should come on top of the half-quantized Majorana Hall effect. Therefore the behaviour of the Hall signal predicted for a pure Kitaev model will be qualitatively modified when visons are thermally excited at finite temperatures. 
Here an important factor is the relative sign of the Majorana Hall effect and the vison Hall effect. In principle, these are independent parameters. We find that this vison hopping amplitude is {\em not} affected by the sign of the Majorana mass gap $\kappa$.
Within our perturbation theory linear in $h$, we find that the sign of the Chern number of the lowest vison band is determined by the flux enclosed when the vison hops along a triangular loop using hopping processes triggered by $h_x$, $h_y$ and $h_z$.
This results in the Chern number $C_V = -\text{sgn}(h_xh_yh_z)$ for the lowest vison band. This has to be compared to the Chern number of the Majorana band \cite{Kitaev06,kasahara2}, $C_m =\text{sgn}(\kappa)$ which leads to  $C_m = \text{sgn}(h_xh_yh_z)$ for a Kitaev model perturbed by $\boldsymbol h=(h_x, h_y,h_z)$ only \cite{Kitaev06}. As the signs are opposite, the vison Hall effects of Majorana fermions and visons is subtractive.

 \begin{figure*}
	\centering
	\begin{tikzpicture}
	\node[anchor=center] (image) at (0,0) {\includegraphics[width=0.45\textwidth]{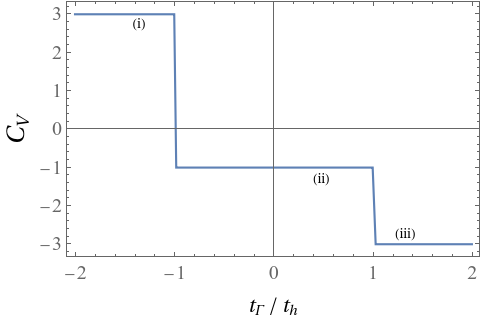}};
	\node[anchor=center] (text) at (-3.8,2.2) {\small{(a)}};
	\end{tikzpicture}
	\begin{tikzpicture}
	\node[anchor=center] (image) at (0,0) {\includegraphics[width=0.45\textwidth]{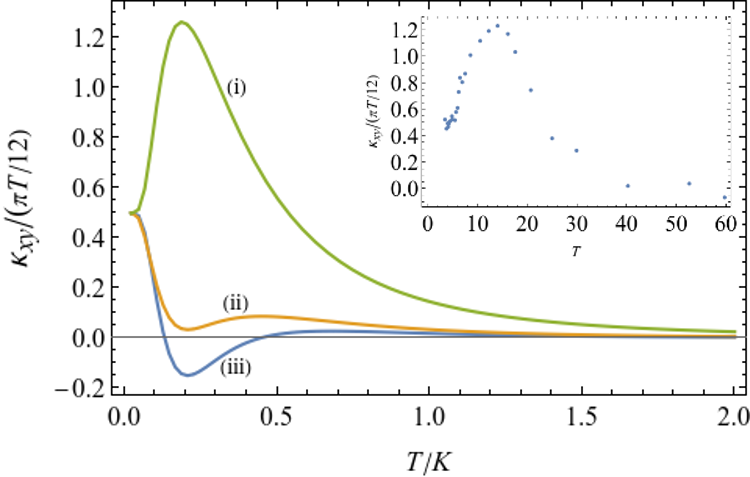}};
	\node[anchor=center] (text) at (-3.5,2.2) {\small{(b)}};
	\end{tikzpicture}
	\caption{(a) Chern number of the lowest vison band in the AFM Kitaev model as a function of the ratio of $t_\Gamma$ and $t_h$. (b) Thermal Hall conductivity of an AFM Kitaev spin liquid with both matter Majorana and vison contribution. Different curves are obtained for lowest Majorana band having Chern number +1 and lowest vison band with Chern number (+3,-1,-3) as marked in subfigure (a). We assume that the vison bands lie within a large Majorana gap $\triangle_m$. The curves in subfigure (b) are calculated with the follwing parameters: (i)$(E^0_v=0.6\,K,\triangle_m = 0.5\,K,t_h=0.05\,K,t_\Gamma=-0.08K)$,(ii) $(E^0_v=0.6\,K,\triangle_m = 0.5\,K,t_h=0.1\,K,t_\Gamma=0.05K)$, (iii)$(E^0_v=0.6\,K,\triangle_m = 0.5\,K,t_h=0.05\,K,t_\Gamma=0.07K)$. Inset: Experimentally obtained $\kappa_{xy}$ for $\alpha$-RuCl$_3$ (reproduced from Ref.\cite{kasahara2}) \label{fig:visonHall}}
\end{figure*}
We find that if the Majorana gap $\kappa$ solely arises at cubic order in the magnetic field i.e, $\kappa \propto h^3$, then the lowest vison band has the Chern number $1$ with the same sign as that of the lowest Majorana band.
As shown in Fig.~\ref{fig:visonHall}.a, the situation changes when one adds the effect of $t_\Gamma$. Depending on the sign and size of $t_\Gamma$,
the Chern number of the lowest vison band takes the values $3$, $1$, or $-3$. Remarkably, the vison band gets a large Chern number $+3$ when $t_\Gamma/t_h <-1$.

Experimentally, one can expect either a characteristic dip or a peak in the Hall signal depending on whether the Chern number of the lowest vison band is negative or positive as shown schematically in Fig. \ref{fig:visonHall}.

Experimentally, in $\alpha$RuCl$_3$ a characteristic peak above a half-integer quantized plateau has been observed in the thermal Hall effect \cite{kasahara1,kasahara2}. This suggests that the system hosts additional chiral excitations on top of the Majorana fermions. As the amplitude of the peak is very large, almost twice the plateau value, the experimental result is consistent with the presence of a gapped excitation with a Chern number larger than 1. It is tempting to associate this feature with a vison Hall effect but this would require that the spin liquid state has the same projective symmetry group as the {\em antiferromagnetic} Kitaev model in an external field. While it has been suggested early on \cite{KeeAfm} that $\alpha$RuCl$_3$ has an antiferromagnetic Kitaev coupling, experimental evidence is in favor of a ferromagnetic Kitaev coupling, see, e.g., Ref.~\cite{tableofcouplings}.

Above, we considered a magnetic field in $(111)$ direction, perpendicular to the plane. When the field is rotated, the sign of the Hall effect (for both the plateau and the peak) in $\alpha$RuCl$_3$ is approximately given by $\text{sgn}(h_xh_yh_z)$ \cite{kasahara2}.
This is consistent with theory as the Majorana mass $\kappa$ (and thus the Majorana Hall effect) is proportional to $h_xh_yh_z$ \cite{Kitaev06} in the $h$ field-perturbed Kitaev model. The Chern number of the vison band arising from $t_h$ only, is determined by sign of the flux enclosed by a vison hopping on a triangle which is also determined by the product $h_xh_yh_z$. As, furthermore, 
 $t_\Gamma \to t_\Gamma^*$ for $\kappa\to -\kappa$, we find that the sign of the Chern number of the vison band jumps within our approximations simultaneously with the sign of the Majorana Chern number.
}

\section{Discussion}
Depending on temperature and the sign of the Kitaev coupling $K$, we find that a vison can either behave as a coherent quasiparticle with very large mobility or as an incoherent excitation with a small mobility. For antiferromagnetic Kitaev coupling interference effects eliminate all leading order vison-tunneling processes.
This immediately explains why the antiferromagnetic Kitaev model is much more robust against perturbations by $\Gamma$ or $h$ than its ferromagnetic counterpart. In the ferromagnetic case the vison gap shrinks for increasing vison hopping, thus triggering a phase transition when the vison gap closes, see App.~\ref{App:compare} for a more detailed analysis and a quantitative comparison to existing numerical studies.

Our theory provides a controlled calculation in the limit of weak perturbations to Kitaev models. As such it cannot be directly applied to materials like  $\alpha$-RuCl$_3$ where at zero magnetic field these perturbations induce magnetic order, thus destroying the spin liquid state. The observation of a half-integer quantized thermal Hall effect in this material \cite{kasahara1,Yamashita2020,kasahara2,bruin2021robustness} at a field of about 10\,T, however, suggests that 
this field-induced phase is adiabatically connected to the physics of a ferromagnetic \cite{proximate,FMKitaev1,FMKitaev2,anisotropy,FMKitaev3} Kitaev model weakly perturbed by a magnetic field. 
Thus it is highly plausible
that this phase also hosts a dynamical gauge field. \new{The fact that the quantized Hall effect has been seen only in few samples \cite{Yamashita2020,taillefer} however, complicates the experimental interpretation.
The presence of vison bands with non-trivial topology can also show up in the thermal Hall effect measurements. We showed that the presence of a $\Gamma$ and a $(111)$ magnetic field perturbation can give rise to vison bands with both positive and negative $(\pm 3$, $-1$) Chern numbers, depending on their relative strength and sign. This in turn could lead to a characteristic peak or a dip on top of the half-quantized Majorana Hall plateau, see Fig. \ref{fig:visonHall}. In parallel to our study, the vison Chern bands were also studied by Chuan Chen and Inti Sodemann Villadiego \cite{Intiafm} using an exact fermion lattice duality.
 }

%

\new{Arguably, one of the most promising routes to detect the dynamics of visons is to study the equilibration dynamics of a perturbed
Kitaev spin liquid. Vison diffusion is essential for equilibration and at low temperatures it is governed by a fully universal diffusion constant $D=\frac{6 \hbar v_m^2}{k_B T}$.  We therefore suggest to search for signatures of vison dynamics in the long-time tails of pump-probe experiments \cite{loosdrecht}. }

\begin{acknowledgments}
We acknowledge useful discussions with Jinhong Park, Simon Trebst, Martin Zirnbauer and, especially, Ciar\'an Hickey. 
We would like to thank especially Chuan Chen, Peng Rao and  Inti Sodemann Villadiego for pointing out a mistake in an early version of the manuscript.
This work was supported by the
Deutsche Forschungsgemeinschaft (DFG) through CRC1238 (Project
No. 277146847, project C02 and C04) and -- under Germany’s Excellence
Strategy -- by the Cluster of Excellence Matter and Light for Quantum
Computing (ML4Q) EXC2004/1 390534769 and by the Bonn-Cologne Graduate School of Physics and Astronomy (BCGS).
\end{acknowledgments}

\appendix

\section{Matrix element calculation}\label{App:me}
\begin{figure}[!h]
	\centering
	\includegraphics[width=0.4\textwidth]{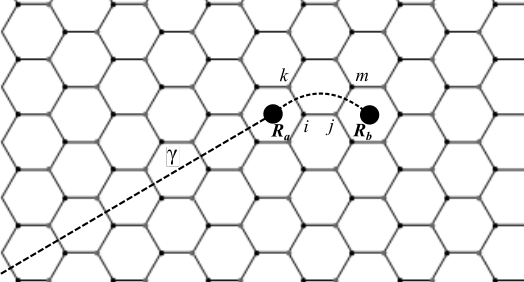}
	\caption{Schematic showing the gauge configurations for the vison states used in the computation of matrix elements (Eq.~\ref{eq:gamma_cal}). In a periodic system, a vison at $R_a$ is created by flipping the $u_{ij}$ variables along the dashed line. The other end of the line carries another vison far separated. \label{fig:gaugeConfig}}
\end{figure}
In this section, we describe the Pfaffian method \cite{Robledo} for calculating hopping matrix elements, Eq.~\eqref{eq:matrixHoppig}, which involve the overlap of different Bogoliubov vacua. The starting point of our analysis is the many-body wave function, Eq.~\eqref{eq:spin_wf} in the main text, of Majorana fermions scattering from a localized vison (or a pair of visons, see below),
\begin{align}
	\ket{\Phi(\boldsymbol{R}_a)} = \hat{P}\prod_{l\in\gamma} \chi^\dagger_l\ket{0_\chi}\ket{M_0(\boldsymbol{R}_a,\mathcal{G}_a)}.\label{eq:wf_app}
\end{align}
The gauge configuration is expressed in terms of the bond fermion wave-functions,
where $\gamma$ is a semi-infinite string of $x$ links flipped by the action of $\chi^\dagger_l$ on the bond fermion vacuum $\ket{0_\chi}$. $\ket{M_0(\boldsymbol{R}_a,\mathcal{G}_a)}$ is the many-body ground state wave function of the matter fermions in the chosen gauge. Note that for $\ket{\Phi_0(\boldsymbol{R}_a)}$ we have the freedom to choose any gauge configuration but the projection operator ensures gauge invariance. It is easy to see that the most convenient choice to relate two vison wave functions located at positions $\boldsymbol{R}_a$ and $\boldsymbol{R}_b$, see Fig.~\ref{fig:gaugeConfig}, is $\ket{\mathcal{G}_b} = \chi^\dagger_{\langle mj \rangle_y}\chi^\dagger_{\langle ik \rangle_x}\ket{\mathcal{G}_a}$. Eliminating the gauge sector by contracting the bond fermions, Eq.~\eqref{eq:matrixHoppig} for $\Gamma$ perturbation becomes
\begin{widetext}
\begin{align}
	\label{eq:gamma_cal}
		t^\Gamma_{ab}  &= \Gamma\bra{M_0(\boldsymbol{R}_a,\mathcal{G}_a)}\bra{\mathcal{G}_a}(b^y_ib^x_jc_ic_j+b^x_ib^y_jc_ic_j\hat{D}_j\hat{D}_i)
		\ket{\mathcal{G}_b}\ket{M_0(\boldsymbol{R}_b,\mathcal{G}_b)}\\ \nonumber
		&= \Gamma\bra{M_0(\boldsymbol{R}_a,\mathcal{G}_a)}(-ic_ic_j-1)\ket{M_0(\boldsymbol{R}_b,\mathcal{G}_b)}
\end{align}
\end{widetext}
Similarly, one can show that for $\Delta H_h$, the matrix element for hopping across the $\langle ij \rangle_z$ link can be written as
\begin{align}
	t^h_{z} = \bra{M_0(\vec R_a,\mathcal{G}_a)}\left(-i+c_ic_j\right)\ket{M_0(\boldsymbol{R}_b,\mathcal{G}_b)}
	\label{eq:h_cal}
	\end{align}
where we used a gauge transformation for the spin operator which is equivalent to rewriting $\sigma^z = -i\sigma^x\sigma^y$. This is possible if the states The positions $R_a$ and $R_b$ are defined in  Fig.~\ref{fig:interference}b.  
We used the following decomposition of the projection operator that relates it to the total fermionic parity (bond and matter fermions) \cite{Loss}.
\begin{align}
	\label{eq:proj_decomp}
	\hat{P} &=  \hat{P}'\frac{(1+\prod_i\hat{D}_i)}{2} = \hat{P}'\frac{1+(-1)^{\theta + N_\chi+N_f}}{2}
\end{align}
where $\theta \in \mathbb{Z}$ is a geometric factor that depends on the lattice boundary conditions, see Ref.~\cite{vojta} for details.
This helps to avoid choosing an unphysical state while evaluating Eq.~\eqref{eq:gamma_cal} and Eq.~\eqref{eq:h_cal} (for finite systems) which would otherwise give zero as $\hat P$ projects away any unphysical state. Hence we choose the gauge configuration ($\mathcal G_a, \mathcal G_b$) such that the ground states are physical by calculating the fermionic parities explicitly using the methods discussed in Refs.~\cite{Loss,vojta}.\\

For our calculation we use periodic boundary conditions with {\em two} visons placed at a large distance. The position of the second vison is always kept fixed (with its position coordinate suppressed in Eq.~\eqref{eq:wf_app}) while the position of the first vison is denoted by $\vec R_a$. To compute the matrix elements, we first diagonalize the Majorana Hamiltonian with a vison at a reference position $\vec R_d$, $\vec R_a$ and $\vec R_b$ using suitable gauge configurations. The corresponding Bogoliubov transformations are of the form
\begin{align}
	\label{eq:ab_trans}
	\begin{pmatrix}
		{X}^{(a)*} && {Y}^{(a)*}\\
		{Y}^{(a)} && {X}^{(a)}\end{pmatrix} 
	\begin{pmatrix}
		f \\ f^\dagger\end{pmatrix} = \begin{pmatrix}
		a\\a^\dagger\end{pmatrix}
\end{align}
for $\vec R_a$ and the $a\leftrightarrow b,d$ for $\vec R_b$ and $\vec R_d$ respectively.
We define a reference vacuum $\ket{\tilde{0}}$ and fermionic operator $d_i$ with $d_i \ket{\tilde{0}}=0$  \cite{robledo2}. Importantly, this state must have the same total fermion parity as the two ground states of our interest and must be physical. One can choose this to be, say the ground state of a third vison position. The Bogoliubov operators $a$, which diagonalize the Kitaev model for a vison located at position $\vec R_a$ can be related to $d$ by unitary matrices (similarly for $\vec R_b$).
\begin{align}
	\begin{pmatrix}
		\mathcal{X}^{(a)*} && \mathcal{Y}^{(a)*}\\
		\mathcal{Y}^{(a)} && \mathcal{X}^{(a)}\end{pmatrix} 
\begin{pmatrix}
	d \\ d^\dagger\end{pmatrix} = \begin{pmatrix}
	a\\a^\dagger\end{pmatrix}
\end{align}
with $\mathcal{X}^{(a)*}  = Y^{(a)}Y^{(d)\dagger}+X^{(a)}X^{(d)\dagger}$ and $\mathcal{Y}^{(a)*}  = Y^{(a)}X^{(d)\dagger}+X^{(a)}Y^{(d)\dagger}$.

\begin{figure}
	\includegraphics[width=0.2 \textwidth]{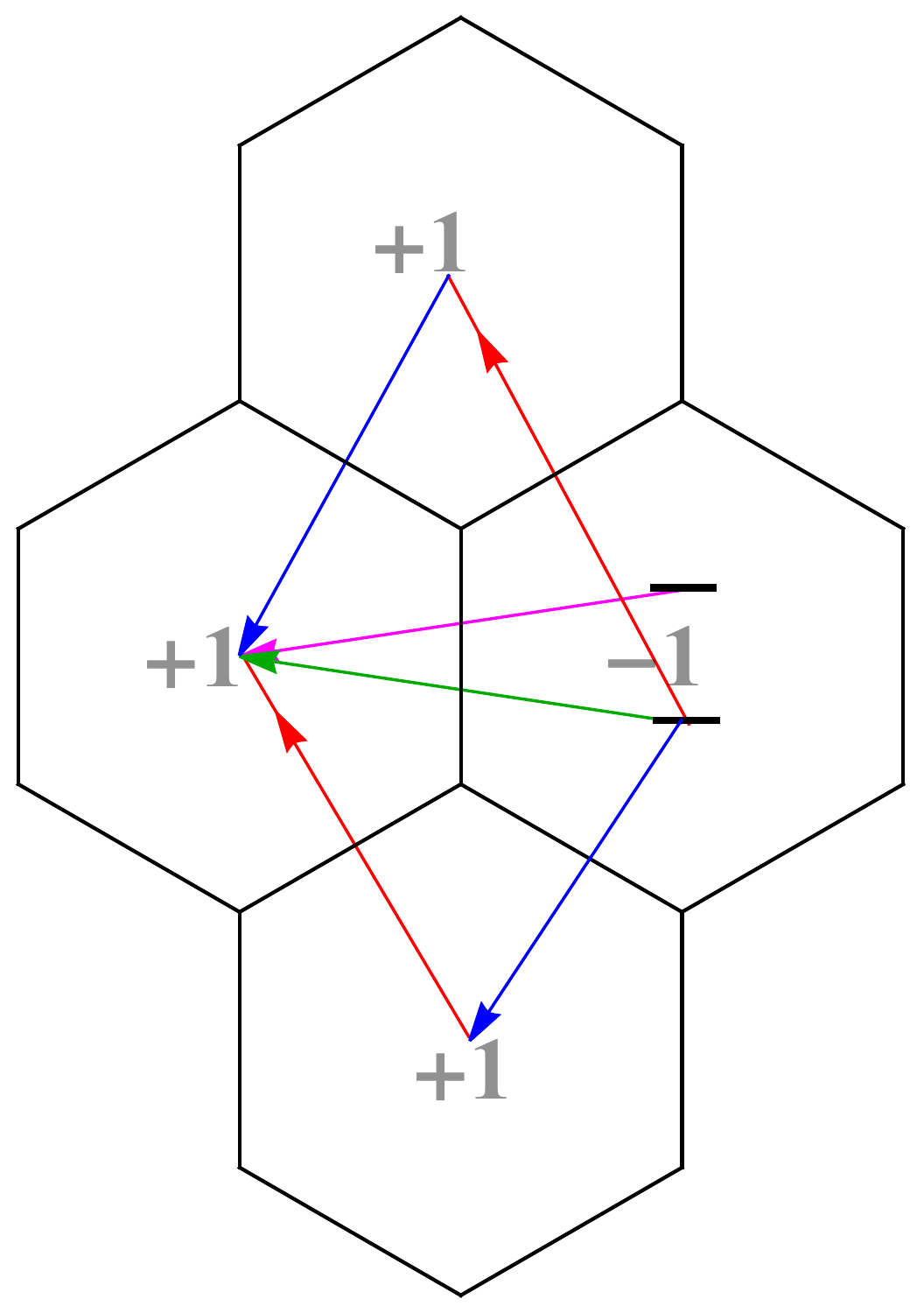}
	\caption{\new{
			The pattern of ground-state fermionic parity is indicated as +1 (odd) or +1 (even) in the plaquette where the vison is located. This is fixed for a given position of the second vison which is placed on the far left (not shown in the figure). For the odd parity case, one needs to calculate the hopping amplitudes to states with a single particle added to the BCS vacuum to stay in the physical Hilbert space. The two levels shown in the odd parity plaquette denote the 1st two levels of the Majorana spectrum.}}\label{fig:parity}
\end{figure}

 We can now express both $\ket{M_0(R_a,\mathcal{G}_a)}$ and $\ket{M_0(\vec R_b,\mathcal{G}_b)}$ in the following Thouless form \cite{Robledo,knolle},
\begin{align}
	\label{eq:thouless}
	\ket{M_0(\boldsymbol{R}_a,\mathcal{G}_a)} =\left|\det(\mathcal{X}^{\left(a\right)})^{\frac{1}{2}}\right|e^{-\frac{1}{2}d^\dagger Z^{(a)}d^\dagger}\ket{\tilde{0}},
\end{align}
with $Z^{(a)} = \left(\mathcal{X}^{(a)^{-1}} \mathcal{Y}^{(a)}\right)^*$.
Matrix elements of the form needed for Eq.~\eqref{eq:gamma_cal} and \eqref{eq:h_cal} can be computed using $d  (d^\dagger)$ operators grouped into $(\tilde{d}_1...\tilde{d}_{2N}) = (d^\dagger_1...d^\dagger_Nd_1...d_N)$. Matrix elements of  $ \tilde{d}_m\tilde{d}_n$ can be computed using a coherent state path integral technique to give
\begin{align}
	\begin{split}
	\bra{M_0(\boldsymbol{R}_a,\mathcal{G}_a)}\tilde{d}_m& \tilde{d}_n\ket{M_0(\boldsymbol{R}_b,\mathcal{G}_b)} \\ &= (-1)^{N(N+1)}\text{Pf} (\mathbb{X})\,\text{Pf}(\mathbb{X}_{\{m,n\}})
	\end{split}
\end{align}
where $\text{Pf}$ denotes the Pfaffian and $\mathbb{X}$ is a $2N \times 2N$ skew-symmetric matrix defined using $Z^{(a)} $ and $Z^{(b)}$,
\begin{align}
	\mathbb X = \begin{pmatrix}-Z^{(b)^*}&&-I\\I && Z^{(a)}\end{pmatrix}.
\end{align}
where $\mathbb{X}_{\{m,n\}}=\begin{pmatrix} 0&&\mathbb{X}_{mn}\\\mathbb{X}_{nm}&&0\end{pmatrix}$ is a $2 \times 2$ matrix.
Pfaffians were computed using the algorithm developed by Wimmer \cite{wimmerPf}.
\new{\subsection{Ground state parity and $h$ induced hopping}
As discussed in the main text, a $(111)$ magnetic field $h$, hops a vison between nearest neighbour plaquettes.
While evaluating such an overlap, it turns out that, for certain relative vison positions, the Bogoliubov vaccum state as defined in Eqn.\ref{eq:thouless} is unphysical since it has an odd fermionic parity. Therefore one should add an extra Boguliubov particle to the vacuum to get the true physical states. So the physical states in the case of odd parity are given by
\begin{align}
	\ket{M^{odd}_l(\boldsymbol{R}_a,\mathcal{G}_a)} =\left|\det(\mathcal{X}^{\left(a\right)})^{\frac{1}{2}}\right|a^\dagger_le^{-\frac{1}{2}d^\dagger Z^{(a)}d^\dagger}\ket{\tilde{0}},
	\end{align}
where $l=0$ gives the physical ground state and $l=1$ gives the first excited state.
This results in a pattern of ground-state parities as illustrated in Fig. \ref{fig:parity}, for a given position of the second vison and a fixed gauge configuration (not shown in the figure). While hopping along across the $y$ bond from a +1 plaquette to -1 plaquette, one therefore has to calculate the following overlaps for $l=0$ and $l=1$.
\begin{widetext}
	\begin{align}
		\label{eq:gamma_cal_odd}
		t^{h(l)}_{z}  &= \Gamma\bra{M^{odd}_l(\boldsymbol{R}_a,\mathcal{G}_a)}\bra{\mathcal{G}_a}(\sigma_i^z+\sigma_j^z)
		\ket{\mathcal{G}_b}\ket{M_0(\boldsymbol{R}_b,\mathcal{G}_b)}\\ \nonumber
		&= \Gamma\bra{M^{odd}_l(\boldsymbol{R}_a,\mathcal{G}_a)}(ib^z_ic_i+ib_j^zc_j)\hat{P}\ket{M_0(\boldsymbol{R}_b,\mathcal{G}_b)}
	\end{align}
\end{widetext}
These can be evaluated using the same Pfaffian method as described in Appendix.\ref{App:me}. Since the true physical ground state for an odd parity state is obtained by filling the lowest energy mode which is the (quasi-)localized Majorana zero mode (MZM), it interacts with the second vison if the localization length of the MZM wavefunction is larger than the distance between the visons. This finite-size effect results in a breakdown of the validity of an isolated vison theory, in the small Majornana gap limit. However, for a $nnn$ hopping as induced by the $\Gamma$ term, the many-body wavefunctions are of the same parity and hence this finite-size effect is absent.}

\section{Scattering from a static vison}\label{App:B}
In this section we briefly review the scattering of low-energy Majorana degrees of freedom from a single, static vison. We will need the result to compute the 
mobility of mobile visons in the next section, App.~\ref{App:mobility}.
At low energies the matter Majoranas, $c$, are described by Dirac equation with velocity $v_m=\sqrt 3 |K|/2$ at momenta $\vec K$ and $\vec K'$. Using the property $c_k=c^\dagger_{-k}$, we can combine the two Majorana cones into one single Dirac cone at $\vec K$ and restrict the momenta to half-Brillouin zone. Expanding around the momentum $\vec K$ one obtains in radial coordinates
\begin{align}
\tilde{H}_{\vec K} = v_m
\begin{pmatrix}
0& ie^{ i\theta}\left(\partial_r + \dfrac{i}{r}\partial_\theta\right)\\
 ie^{- i\theta}\left(\partial_r - \dfrac{i}{r}\partial_\theta\right)&0 
\end{pmatrix}
\label{eq:dirac_ham}	
\end{align}
The vison is described as a point-like magnetic flux with flux $\pi$ located at the origin of the coordinate system. We use a gauge where the presence of the flux can be absorbed into antiperodic boundary conditions in $\theta$ direction, $\psi(\theta)=-\psi(\theta+2 \pi)$. 
 This is equivalent to a singular gauge often used in vortex scattering problems \cite{tesanovic}.
The scattering solutions can be obtained by solving a second order Bessel differential equation
\begin{align}
	\Tilde{\psi}_{s,l,k}(r) =\left\{ \begin{array}{ll}
	& \sqrt{\frac{k}{4\pi}}\begin{pmatrix}
		s^{\frac{1}{2}}J_{-l+\frac{1}{2}}(kr) e^{i(l-\frac{1}{2})\theta}\\
		-s^{-\frac{1}{2}}iJ_{-l-\frac{1}{2}}(kr) e^{i(l+\frac{1}{2})\theta}
	\end{pmatrix}, \quad l\leq 0 \\
	& \sqrt{\frac{k}{4\pi}}\begin{pmatrix}
		s^{\frac{1}{2}}J_{l-\frac{1}{2}}(kr) e^{i(l-\frac{1}{2})\theta}\\
		s^{-\frac{1}{2}}iJ_{l+\frac{1}{2}}(kr) e^{i(l+\frac{1}{2})\theta}
	\end{pmatrix}, \quad l> 0
\end{array}\right. \label{eq:scatt_wf}
\end{align}
where $s=\pm 1$ labels the positive and negative energy states respectively.\\
The case of $l=0$ is special. The wave function weakly diverges at the origin as $r^{-\frac{1}{2}}$ and thus is a quasi-localized state \cite{perkins}.
The well-known scattering cross-section can be obtained as \cite{AB,durst}
\begin{equation}
\frac{d\sigma}{d\varphi} = \frac{1}{2\pi k \sin^2{\varphi}}, \qquad \varphi \ne 0 \label{eq:sigma}
\end{equation}
where $\varphi$ is the angle between incoming and outgoing beam. 
\section{Mobility of a vison}\label{App:mobility}
To discuss the mobility of a a single mobile vison, we use the language of a Boltzmann equation for the momentum distribution function $f_{\vec p}=f^0_{\vec p}+\delta f_\vec{p}$ of the vison. 
We argue that the Boltzmann equation (and further approximations to the Boltzmann equation discussed below) becomes exact
in the limit of low $T$. In this limit the density of visons is exponentially small and thus we can focus on the properties of a single vison ignoring vison-vison interactions and also effects like a finite lifetime of Majorana states due to vison-Majorana interactions. We will furthermore use below that visons are much slower than Majorana fermions. Also the density of Majorana excitations, $n_m$,  vanishes as $n_m \sim T^2$ for low $T$. A semiclassical approximation is valid if the mean-free path $\xi_v$ of the vison is large compared to its wavelength $\lambda_v$. Here it is important to take into account the diverging cross sections, Eq.~\eqref{eq:sigma}. We can estimate $\xi_v$ from $\xi_v \sigma n_m =1$. Using $k \sim T$ for Majorana fermions, we obtain $\xi_v \sim \frac{1}{T} \gg \lambda_v \sim \frac{1}{\sqrt{T}}$, justifying the use of a semiclassical approximation \cite{RammerSmith1986}. 
The low density of Majorana fermions at low $T$ also justifies that we neglect  Majorana-Majorana interactions which is an irrelevant perturbation in the RG sense.

In the presence of an external force $\vec F$ acting on the vison, the linearized Boltzmann equation reads
\begin{align}
	\vec F \cdot \vec v^v_\vec{p} \frac{\partial f^0_{\vec p}}{\partial E^v_{\vec{p}}} = 
	 \int \tilde M_{\boldsymbol p \boldsymbol p'}\delta f_{\boldsymbol p'} \frac{d^2 p'}{(2 \pi)^2}
\end{align}
Here the equilibrium distribution function of the gapped vison, $f^0_{\vec p}= c \, e^{-\beta  E^v_\vec{p}}$, is given by a Boltzmann distribution as we work in the low-density limit and $c$ is a normalization constant which will drop out in the final result. $E^v_\vec{p}$ is the dispersion of the vison,
 $\vec v^v_\vec{p}=\partial E^v_\vec{p}/\partial \vec p $ its velocity. 
The scattering rate from momentum $\vec p'$ to momentum $\vec p$ 
 is determined from
 \begin{align}
	M_{\boldsymbol p,\boldsymbol p'} &= \int \frac{d^2k}{(2\pi)^2}\frac{d^2k'}{(2\pi)^2} W^\vec{p}_{\boldsymbol k,\boldsymbol k'}\ n^0_k(1-n^0_{k'}) \nonumber\\ & ´\quad
	\delta(\boldsymbol k+\boldsymbol p-\boldsymbol k'-\boldsymbol p')\delta(\epsilon_{\vec k}+E^v_{\vec{p}'}-\epsilon_{\vec k'}-E^v_{\vec p}).\label{eq:Mexact}
\end{align}
with $\tilde M_{\boldsymbol p \boldsymbol p'}=M_{\boldsymbol p \boldsymbol p'}-\delta(\vec p-\vec p') \int M_{\vec p' \vec p} d^2p'$, where the second term describes the out-scattering from $\vec p$ to an arbitrary momentum $\vec p'$.
	As we consider a single vison embedded by many thermally excited Majorana modes, we can assume that the latter stay in equilibrium. Thus $n^0_k$ is the Fermi distribution function in equilibrium. We consider the case where the Majorana dispersion arises from a small $\Gamma$ term and we focus on the limit $T\ll K$. Thus, we can approximate the Majorana dispersion by 
	$\epsilon_\vec{k} \approx v_m |\vec k|$. The transition rates $W^\vec{p}_{\boldsymbol k,\boldsymbol k'}$ are discussed below.

We use the ansatz  $\delta f_{\boldsymbol p} = \frac{\partial f_0}{\partial E^v_{\vec p}} \phi_{\boldsymbol p}$, where $\phi_{\boldsymbol p}$ is a smooth function in momentum and obtain
\begin{align}
	\begin{split}
		\vec F \cdot \vec v^v_\vec{p} =&\int \tilde M_{\boldsymbol p \boldsymbol p'}e^{\beta(E^v_{\vec p}-E^v_{\vec p'})} \phi_{\boldsymbol p'} \frac{d^2 p'}{(2\pi)^2}
	\end{split}
\end{align}
A substantial simplification of this matrix equation occurs because (i) the vison velocities are much smaller than Majorana velocities and (ii) due to $T \ll K$ the typical momenta of the Majorana modes, $\sim k_B T/v_m$, are small. Due to energy and momentum conservation, therefore the typical vison momentum transfer, $|\vec p - \vec p'| \sim k_B T/v_m$, is also small. Therefore 
one can expand the smoothly varying function $\phi_{\boldsymbol p'}$ and also $E^v_\vec{p}-E_{\vec{p}'}$ in the momentum difference retaining only the leading order terms. A similar approach has, for example, been used to describe the relaxation of high-energy quasiparticle in d-wave superconductors \cite{howell2004}. Thus, we arrive at
\begin{align}
	\boldsymbol v^v_{\boldsymbol{p}}\cdot \boldsymbol F \approx &\int \tilde M_{\boldsymbol p \boldsymbol p'}e^{-\beta \vec{v}^v_\vec{p} (\vec{p}'-\vec{p})} \Bigl(\phi_{\boldsymbol p}+(\boldsymbol p'-\boldsymbol p)\cdot \nabla_{\boldsymbol p}\phi_{\boldsymbol p} )\nonumber \\
	&\quad+\frac{(p'_i-p_i)(p_j'-p_j)}{2}\partial_{p_i} \partial_{p_j} \phi_{\boldsymbol p}\Bigr)\,\frac{d^2 p'}{(2 \pi)^2}
	\label{eq:taylor}
\end{align}
The zeroth  order terms vanish exactly due to the outscattering term in $\tilde M$. In the limit of vanishing vison bandwidth, $\vec v^v_\vec{p}\to 0$, also the second term vanishes as $\tilde M_{\boldsymbol p,\boldsymbol p'}$ is only a function of $|\vec p - \vec{p}'|$ in this case. Therefore, we have to compute this term to linear order in $\vec v^v_\vec{p}$, while this is not necessary for the second-order term. Thus we arrive at the following drift-diffusion equation in momentum space
\begin{align}
	\partial_t \phi_\vec{p}+\boldsymbol v^v_\vec{p}\cdot \boldsymbol F \approx D_p  \nabla^2_{\vec p} \phi_{\vec p}+\gamma \, \vec v^v_\vec{p}\cdot \vec{\nabla}_\vec{p}  \phi_{\boldsymbol p}\label{eq:driftdiffusion0}
\end{align}
with yet undetermined prefactors $D_p$ and $\gamma$. The ratio of $\gamma$ and $D_p$ can be determined without any microscopic calculation by demanding that Eq.~\eqref{eq:driftdiffusion0} obeys particle number conservation for arbitrary $\phi_\vec{p}$. From this condition, we derive $\gamma=-D_p/T$ and obtain 
\begin{align}
	\partial_t \phi_\vec{p}+\boldsymbol v^v_\vec{p}\cdot \boldsymbol F \approx D_p \left( \nabla^2_{\vec p} \phi_{\vec p}-\frac{1}{T} \, \vec v^v_\vec{p}\cdot \vec{\nabla}_\vec{p}  \phi_{\boldsymbol p}\right)\label{eq:driftdiffusion}
\end{align}
or, after rewriting the result in terms of the vison distribution function $f_\vec{p}$ we obtain the equivalent equation
	\begin{align}
	\partial_t f_\vec{p}+\boldsymbol v^v_\vec{p}\cdot \boldsymbol F \frac{d f^0}{d E^v_\vec{p}} \approx D_p \left( \nabla^2_{\vec p} f_{\vec p}+\frac{1}{T} \, \vec \nabla_\vec{p} \left( \vec v^v_\vec{p}f_{\vec p}\right)\right)\label{eq:driftdiffusionf}.
\end{align}
The two equations \eqref{eq:driftdiffusion} and \eqref{eq:driftdiffusionf} describe the Brownian motion of the vison. 
There is a frictional  force proportional to $-\vec v^v_\vec{p}$ which slows the vison down. This dissipation is necessarily accompanied by fluctuations: random forces due to vison-Majorana scattering lead to a diffusion in momentum space.


Due to the momentum dependence of the drift term, Eq.~\eqref{eq:driftdiffusion} cannot be solved analytically but we obtain a numerical solution 
by Fourier transformation followed by a matrix inversion. In the low-$T$ limit it is important to take a sufficient number of Fourier components into account as $\Phi_\vec{p}$ develops features with a width $\sim \sqrt{T}$.

Analytically, one can solve the the drift-diffusion equation for $T \gg W_v$ simply by ignoring the drift term proportional to $v^v_\vec{p}$ and by integrating the dispersion twice maintaining periodic boundary conditions. In the low-$T$ limit, $T \ll W_v$, the stationary equation is approximately solved by
$\phi_\vec{p} = -\frac{T}{D_p} \vec F \cdot \vec p$. The periodicity of $\phi_\vec{p}$ is thereby restored by a jump of the distribution function far away from the band minimum close to points where $\vec F \cdot v^v_\vec{p}$ vanishes.

The mobility $\mu$ of the vison is computed from
\begin{align}
\langle \boldsymbol v^v_\vec{p} \rangle &= \mu \,{\boldsymbol F}\\
	\langle \vec v^v_\vec{p}\rangle &= \frac{1}{N_v}\int \frac{d^2p}{(2 \pi)^2} \boldsymbol v^v_{\boldsymbol p} \frac{\partial f_0}{\partial E^v_{\vec p}} \phi_{\boldsymbol p} \nonumber
\end{align}
with $N_v=\int f^0_\vec{p} \frac{d^2p}{(2 \pi)^2}$.

We can now use the above described asymptotic solutions for $\phi_\vec{p}$ to calculate analytically the asymptotic behavior of the mobility. We obtain
	 \begin{align}
	 \mu \approx  \left\{\begin{array}{ll}\frac{3 t^2}{D_p T} & \text{for }\ T\gg W_v \\[1mm]
	 \frac{ T}{D_p} & \text{for }\ T\ll W_v
	 \end{array} \right. \label{eq:muAsym}
	 \end{align}
	 where $t$ is the hopping matrix element of the vison.
	 
	 The remaining task is to calculate the temperature dependence of the diffusion constant in momentum space, $D_p$.  By definition $D_p$ is independent of the vison dispersion, therefore its $T$ dependence is a simple power law in this low $T$ regime. This can be obtained in the following way. 
	 A two-dimensional Dirac equation has a linear density of states and therefore the density $n_m$ of thermally excited Majorana fermions is proportional to $T^2/v_m^2$, where $v_m$ is the velocity. The diffusion constant in momentum space is obtained from $(\delta k)^2/\tau$, where $\delta k \sim T/v_m$ is the typical momentum transfer in a scattering event. The scattering time is estimated from $\sigma v_m \tau n_m \sim 1$, where $\sigma$ is the transport scattering cross section which scales with  $1/k$, Eq.~\eqref{eq:sigma}, resulting in an extra factor $v_m/T$, and thus $1/\tau \sim T$.
 Combining these factors one obtains
	 \begin{align}
	 D_p  \sim \frac{T^3}{v_m^2}.\label{eq:Dasym}
	 \end{align}
	
To obtain the correct prefactors, one has to express the transition matrix $W_{\vec{k},\vec{k'}}$ in Eq.~\eqref{eq:Mexact} by the differential cross section for vison-Majorana scattering which is given in Eq.~\eqref{eq:sigma}.
The two quantities are related by \cite{reif}
\begin{align}
	\begin{split}
	\frac{d^2p'}{(2\pi)^2}\frac{d^2k'}{(2\pi)^2}W_{\boldsymbol k,\boldsymbol k'}(2\pi)^2\delta(\boldsymbol k+\boldsymbol p-&\boldsymbol k'-\boldsymbol p')2\pi\delta(\epsilon_{\boldsymbol k}-\epsilon_{\boldsymbol k'}) \\&\approx v_m d\theta_{\boldsymbol k,\boldsymbol k'}\frac{d\sigma(k,\theta_{\boldsymbol k,\boldsymbol k'})}{d\theta_{\boldsymbol k,\boldsymbol k'}}
	\end{split}
\end{align}
This gives, using Eq.~\eqref{eq:taylor}
\begin{widetext}
\begin{align}
	\begin{split}
		D_p 
		=v_m\int \frac{d^2k}{(2\pi)^2}d\theta_{\boldsymbol k,\boldsymbol k'}k^2 \left(1-\cos{\theta_{\boldsymbol k,\boldsymbol k'}}\right)\frac{d\sigma(k,\theta_{\boldsymbol k,\boldsymbol k'})}{d\theta_{\boldsymbol k,\boldsymbol k'}}n(\epsilon_k)(1-n_{\epsilon_{k'}})
		=\frac{T^3}{6 v_m^2}
	\end{split}
\end{align}
\end{widetext}
This fixes the prefactor in Eq.~\eqref{eq:Dasym} in the limit where the vison mass is large. Thus it allows to compute analytically the
exact mobility of the vison both in the low- and high-temperature regime using Eq.~\eqref{eq:muAsym}.

\section{Assisted hopping rate}
\label{App:assisted}
\begin{figure}
	\centering
	\includegraphics[width=0.34\textwidth]{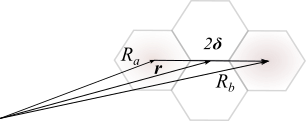}
	\caption{Position vectors of vison used in the calculation of the assisted hopping rate due to a $\Gamma$ perturbation. $\vec r$ is the position vector of the unit cell chosen to be a $z$ bond. \label{fig:assisted}}
\end{figure}

In this section we calculate the mobility in the antiferromagnetic Kitaev model
perturbed by $\Gamma$, similar results apply for a perturbation by a magnetic  field, see below. In this section we use $\vec r$ to label unit cells and $A$ and $B$ to refer to the atom on sublattice $A$ and $B$ within the unit cell.


Consider $\Delta H_\Gamma = \Gamma (\sigma^x_{\vec r,A}\sigma^y_{\vec r,B}+\sigma^y_{\vec r,A}\sigma^x_{\vec r,B})$ with $\vec r$ being the coordinate of the center of the $z$-bond. This term induces a hopping of a vison along a $z$ bond as shown in the Fig.~\ref{fig:assisted}. $\Delta H_\Gamma$ can be written as
\begin{equation}\label{eq:GammaHop}
	\Delta H_\Gamma = \Gamma \left[b^x_{\vec r,A}b^y_{\vec r,B}\left(c^A_{\vec r}-ic^B_{\vec r}\right)\left(c^A_{\vec r}+ic^B_{\vec r}\right)\right]
\end{equation}
where we fixed $ib^z_{\vec{r},A}b^z_{\vec{r},B}=1$ for the two single vison states.
The $b^{x/y}_{\vec r,A/B}$ operators realize the hopping of a bare vison and thus can be simply contracted in the matrix element calculation  as we did in Appendix~\ref{App:me}, see Eq.~\eqref{eq:gamma_cal}. The remaining terms affect the matter Majorana sector which we will treat in the low-energy long-wavelength approximation by replacing the $c$ operators with their continuum fields.
\begin{equation}
c^{A}_{\boldsymbol r}  = \int d^2 r' w(\boldsymbol{r'})e^{\pm i\frac{\pi}{4}}\psi_{A}(\boldsymbol{r'}+\boldsymbol r)+h.c
\end{equation}
 $w(\vec r)$ is a ``Wannier function'' defining an effective cut-off of the low-energy theory. The position of the unit cell is $\vec r = \boldsymbol R_a+\boldsymbol \delta=\boldsymbol R_b-\boldsymbol \delta$ which means that the vison hops by the vector $2 \vec \delta$, see Fig.~\ref{fig:assisted}.
 
 We have shown that the ground-state matrix elements vanish for antiferromagnetic Kitaev coupling. Therefore, we now consider  initial and final states, with a single fermionic excitation above the ground state, which we denote by $\ket{M_n(\vec R_a)}$.  Here $n=\{s,l,k\}$ labels the eigenstates with quantum numbers $s=\pm$ labels particle/hole, $l\in \mathbb Z$ the angular momentum, and energy $\epsilon(k) = v_m k$. Those states will dominate in the low-$T$ limit when the density of thermally excited Majorana states is low.
Thus we need to compute for Eq.~(\ref{eq:goldenRule}) the following matrix elements
\begin{equation}
	\tilde{w}^{ab}(m;n) = \bra{M_n(\vec R_a)}\left(c^A_{\vec r}-ic^B_{\vec r}\right)\left(c^A_{\vec r}+ic^B_{\vec r} \right)\ket{M_m(\vec R_b)}.
	\label{eq:goldenRule1}
\end{equation}

In the continuum theory, we implement the $\pi$ flux carried by a vison as a branch cut that imposes anti-periodic boundary conditions for the Majorana wavefunctions, see App.~\ref{App:B}. As a next step, we expand the field operators in eigenstates of the scattering problem
\begin{align}
	\psi_{A/B}(\boldsymbol{r}-\vec R_{a}) =&\\
	&\hspace{-1cm} \sum_l\int \frac{dk}{2\pi}\sqrt{\pi k}\left(
	a_{+,k,l}-ia_{-,k,l}\right) \left(f^{A/B}_{l,k}(\boldsymbol{r}-\vec R_{a})\right)^* \nonumber 
\end{align}
Here $a_{+,k,l} (a_{-,k,l})$ denote the eigen-modes with $\epsilon_k>0$ $(\epsilon_k<0)$.
\begin{align}
		f^A_{l,k}(\boldsymbol{r}) &= \begin{cases} 
		J_{-l+\frac{1}{2}}e^{l-\frac{1}{2}\theta} e^{iK\cdot r} & l\leq 0 \\
		J_{l-\frac{1}{2}}e^{l-\frac{1}{2}\theta} e^{iK\cdot r} & l > 0 \\
	\end{cases} \nonumber \\
	f^B_{l,k}(\boldsymbol{r}) &= \begin{cases} 
		J_{-l-\frac{1}{2}}e^{l+\frac{1}{2}\theta} e^{iK\cdot r} & l\leq 0 \\
		J_{l+\frac{1}{2}}e^{l+\frac{1}{2}\theta} e^{iK\cdot r} & l > 0 \\
	\end{cases}\label{eq:bessel}
	\end{align}
Note that the low-energy wavefunctions are half-integer Bessel functions naturally arising in vortex-scattering problems \cite{durst,sachdevVortex}.
One can now define particle and hole operators w.r.t the filled Fermi sea.
\begin{align}
	&A^\dagger_+ \equiv a^\dagger_+, \quad A^\dagger_- \equiv a_- \quad \text{with} \quad A_\pm \ket{M_0(\vec R_a)} = 0
\end{align}
Similarly, we denote by $B$ the corresponding operators using scattering states with a vison centered at position $\vec R_b$. 
Expansion of the matrix element, Eq.~\eqref{eq:goldenRule1}, results in a sum of various scattering events $\sim$ $A^\dagger B$, $AB^\dagger$, $A^\dagger B^\dagger $ and $AB$. For a hopping from $\vec R_b$ to $\vec R_a$, we focus on the contribution from terms of the form $A^\dagger B$. They describe processes where both initial and final states contain a single excited Majorana particle. 

In contrast, the term $A B^\dagger$, for example, applied to an initial and finial states with a single excitations can be interpreted as the overlap of vison states with two excitations each. We expect that those give only subleading contributions at low $T$ and focus instead on the $A^\dagger B$ term which is also much easier to compute.

The total transition/hopping rate for a given initial state $n_0=\{s_0,k_0,l_0\}$ denoted by $W^{ab}(s_0,k_0,l_0)=\sum_{s,k,l} \tilde |w^{ab}(s0,k_0,l_0;s,k,l)|^2$ is given by
\begin{widetext}
\begin{equation}
W^{ab}(s_0,k_0,l_0)\approx	\Gamma^2
\left|\bra{M_0(\vec R_a)}\ket{M_0(\vec R_b)}\right|^2  \left(S_{s_0+}(k_0,l_0)+S_{s_0-}(k_0,l_0)\right)
\end{equation}
where the overlap of the ground-state wave functions $\bra{M_0(\vec R_a)}\ket{M_0(\vec R_b)}$ is calculated numerically for a finite size system. 
For a particle excitation in the inital state, $s_0=+$, we obtain
	\begin{align}\label{eq:Spp}
			S_{++}(k_0,l_0) &= \frac{2\pi}{v_m}\sum_{l}\int \frac{dk}{2\pi}\Biggl| \sum_{l_1,l_2} \int d^2r_1d^2r_2 w(\boldsymbol r_1-\boldsymbol \delta)w(\boldsymbol r_2+\boldsymbol \delta) \\
			& \hspace{3cm} \int \frac{dk_1dk_2}{(2\pi)^2}\pi \sqrt{k_1 k_2}\left[\eta^{+}_{k_1,l_1}(\vec{r}_1)\eta^{-*}_{l_2,k_2}(\boldsymbol{r}_2)\right](2\pi)^2\delta(k_0-k_1)\delta(k-k_2)\delta_{l_0,l_1}\delta_{l,l_2}\Biggr| ^2\delta(k_0-k) \nonumber 
\end{align}%
\end{widetext}
where we introduce variables
\begin{align}
	\eta^{\pm^ *}_{k,l}(\boldsymbol r) = e^{i\frac{\pi}{4}}f^{A^*}_{k,l}(\boldsymbol r)\pm e^{-i\frac{\pi}{4}}f^{B^*}_{k,l}(\boldsymbol r).
\end{align}
To obtain $S_{+-}$ one simply has to replace $\eta^-$ by $\eta^+$ in Eq.~\eqref{eq:Spp}.
Substituting the low energy solutions for $f_{k,l}(\vec r)$ from Eq.~\eqref{eq:bessel}, the matrix elements effectively become products of half-integer Bessel functions whose arguments are shifted by the vison separation $\vec 2\delta$. We can also simply replace the  Wannier functions by delta functions for long-wavelength incoming Majorana excitations.
Observing that the leading contribution for $k_0\delta \ll 1$ comes from the $l=0$ state, we get
	\begin{align}
			S_{++}(k_0,l_0)\approx& 
			\frac{\Omega_0^2\pi^2}{v_m}k_0^2\left|\eta^+_{k_0,l_0}(-\boldsymbol \delta)\right|^2\left(k_0\delta+\frac{1}{k_0\delta}\right)
\nonumber \\		
	S_{+-}(k_0,l_0) \approx& \frac{\Omega_0^2\pi^2}{v_m}k_0^2\left|i\eta^+_{k_0,l_0}(-\boldsymbol \delta)\right|^2\left(k_0\delta+\frac{1}{k_0\delta}\right)
\nonumber \\		
	S_{-+}(k_0,l_0) \approx& \frac{\Omega_0^2\pi^2}{v_m}k_0^2\left|-i\eta^{-*}_{k_0,l_0}(-\boldsymbol \delta)\right|^2\left(k_0\delta+\frac{1}{k_0\delta}\right)
\nonumber \\		
			S_{--}(k_0,l_0) \approx& \frac{\Omega_0^2\pi^2}{v_m}k_0^2\left|\eta^{-*}_{k_0,l_0}(-\boldsymbol \delta)\right|^2\left(k_0\delta+\frac{1}{k_0\delta}\right),
	\end{align}
where $\Omega_0$ is the unit cell area. 
The incoherent hopping rate is obtained using the Fermi distribution $n_{k_0,l_0}$ to sum over the initial states.
\begin{align}
	\begin{split}
		W^{ab} \approx& \int \frac{dk_0}{2\pi}\sum_{l_0}n_{k_0,l_0} \Tilde{W}^{(1)}(k_0,l_0)\\
		=& \frac{0.75\Gamma^2\Omega_0^2\pi^3}{32\beta\delta^2 v_m^2}\int du\frac{1}{1+e^u} = \frac{0.39\pi^3a^2\Gamma^2T}{32 v_m^2}.
	\end{split}
\label{eq:linearT}
\end{align}
The result obtained above for a system perturbed by $\Gamma$ can easily be generalized to the case where the perturbation arises from a magnetic field.
In this case the perturbation can be written as
\begin{equation}
\Delta H_h = i h \left[b^x_{\vec r,A}b^y_{\vec r,A}\left(c^A_{\vec r}-ic^B_{\vec r,B}\right)\left(c^A_{\vec r}+ic^B_{\vec r,B}\right)\right],
\end{equation}
where $R_a$ and $R_b$ are nearest neighbour plaquettes as shown in Fig.~\ref{fig:interference}b. The contribution from the $c$ Majoranas is identical to the one in Eq.~\eqref{eq:GammaHop} and thus we obtain the same transition rates with $\Gamma^2$ replaced by $3 h^2$ where the factor $3$ arises because  $ \delta \rightarrow \delta/\sqrt{3}$ due to the smaller hopping distance of the vison in the magnetic-field case.

\section{Heisenberg interaction}
\label{App:heisenberg}
\begin{figure}[!h]
	\centering
	\subfloat[\qquad $\sigma^z_3\sigma^z_2\sigma^x_2\sigma^x_1$\label{subfig:a}]{%
		\includegraphics[width=0.4\columnwidth]{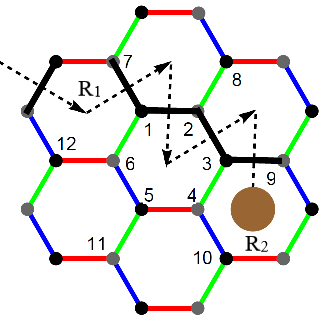}%
	}
	\hfill
	\subfloat[\qquad$\sigma^x_4\sigma^x_5\sigma^z_5\sigma^z_6$\label{subfig:b}]{%
		\includegraphics[width=0.4\columnwidth]{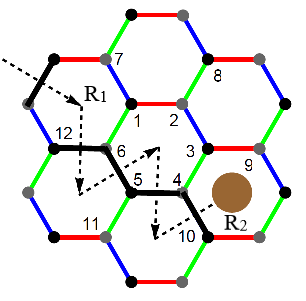}%
	}

	\subfloat[\qquad $\sigma^z_3\sigma^z_2\sigma^y_2\sigma^y_1$\label{subfig:c}]{%
		\includegraphics[width=0.4\columnwidth]{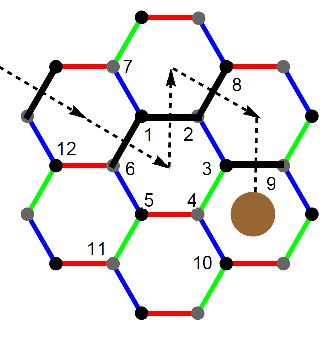}%
	}\hfill
	\subfloat[\qquad $\sigma^x_4\sigma^x_5\sigma^y_5\sigma^y_6$\label{subfig:d}]{%
	\includegraphics[width=0.4\columnwidth]{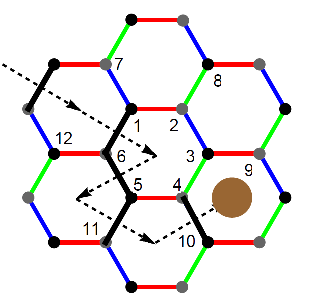}
	}

	\subfloat[\qquad $\sigma^y_3\sigma^y_2\sigma^x_2\sigma^x_1$\label{subfig:e}]{%
	\includegraphics[width=0.4\columnwidth]{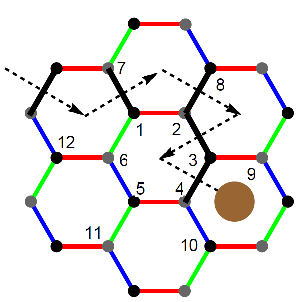}%
	}
	\hfill
	\subfloat[\qquad $\sigma^y_4\sigma^y_5\sigma^z_5\sigma^z_6$\label{subfig:f}]{%
		\includegraphics[width=0.4\columnwidth]{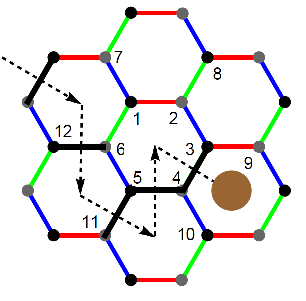}%
	}
	
	\subfloat[\qquad $\sigma^y_4\sigma^y_5\sigma^y_5\sigma^y_6$\label{subfig:g}]{%
		\includegraphics[width=0.4\columnwidth]{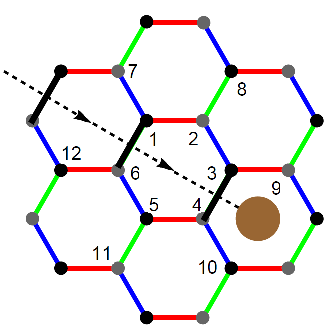}%
	}\hfill
	\subfloat[\qquad $\sigma^y_1\sigma^y_2\sigma^y_2\sigma^y_3$\label{subfig:h}]{%
		\includegraphics[width=0.4\columnwidth]{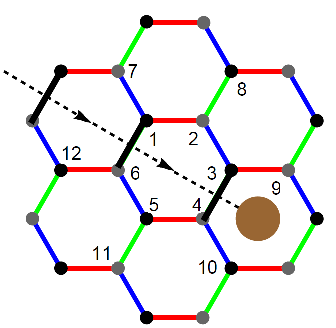}
	}
\caption{Eight single-vison hopping processes ($\vec R_1 \rightarrow \vec R_2$) that pairwise interfere destructively. The dashed arrows pass though the bonds that are flipped (in black), and does not imply a multistep process.	\label{fig:heisenberg}}

\end{figure}

In this section we argue that the single-vison hopping processes induced by the Heisenberg term at order $J^2$ interfere destructively. We consider a hopping across two $y$ links as shown in Fig~\ref{fig:heisenberg}. Let us denote the hopping induced by processes depicted on the left and right side of Fig~\ref{fig:heisenberg} by $t_L$ and $t_R$. A mirror symmetry maps the processes onto each other. We now repeat the argument used in the main text to discuss the interference of hopping processes induced by $\Gamma$ or $h$. By symmetry
$t_L= \pm t_R$ and the sign will decide whether there is a destructive interference, $t_L+t_R=0$, or a constructive interference $t_L+t_R=2 t_L$ of the two terms.

To determine the sign, we analyze a simplified question and consider the sign of
\begin{align}
	\label{eq:var_a}
			\tilde t_{L/R} &= \bra{\Phi^0(\vec R_1)}(\Delta H_J \Delta H_J)_{L/R} \ket{\Phi^0(\vec R_2)},
\end{align}
where we denote by $(\Delta H_J \Delta H_J)_{L/R}$ those terms which contribute to the processes on the left/right side of Fig.~\ref{fig:heisenberg} (written below each figure). Note that $\tilde t_{L/R} \neq t_{L/R}$ but the two quantities are expected to have the same symmetry properties.

To map an $L$ process to a $R$ process we need the information on the flux configuration. The central plaquette in all diagrams in 
Fig.~\ref{fig:heisenberg} does not carry any flux in the initial and final state. The plaquette operator $\hat{W} = \sigma^x_1\sigma^y_2\sigma^z_3\sigma^x_4\sigma^y_5\sigma^z_6$ has eigenvalue $+1$ ($-1$) in the absence (presence) of a flux \cite{Kitaev06}.
Thus, 
\begin{align}
\ket{\Phi^0(\vec R_2)} =\hat{W} \ket{\Phi^0(\vec R_2)}=\sigma^x_1\sigma^y_2\sigma^z_3\sigma^x_4\sigma^y_5\sigma^z_6 \ket{\Phi^0(\vec R_2)}
\end{align}
Using this formula and the algebra of Pauli operators it is straightforward to show that 
\begin{align}
		\bra{\Phi^0(\vec R_1)}\sigma^x_1\sigma^x_2\sigma^z_2&\sigma^z_3\ket{\Phi^0(\vec R_2)} \nonumber \\=&     -\bra{\Phi^0(\vec R_1)}\sigma^z_6\sigma^z_5\sigma^x_5\sigma^x_4\ket{\Phi^0(\vec R_2)}
\end{align}
Therefore the processes shown in Fig.~\ref{fig:heisenberg}a and \ref{fig:heisenberg}b contribute with opposite sign. 

A straightforward extension of this argument is not possible for all the other processes shown in Fig.~\ref{fig:heisenberg}. But a direct evaluation of $\tilde t_L$ and $\tilde t_R$  in a finite size system using the methods from App.~\ref{App:me} reveals that 
\begin{align}
\tilde t_L = - \tilde t_R.
\end{align}
We therefore expect that $t_L=-t_R$ and  processes to order $J^2$ thus cancel by an interference effect independent of the sign of the Kitaev coupling.

A weak Heisenberg coupling is hence expected to contribute only to order $J^4$ to the dispersion of single visons (as $J^3$ terms map a single vison to either 3 or 5 visons). Pairs of visons, however, can even hop by processes linear in $J$ as has been shown in Ref.~\cite{Batista}.

\section{Thermal Hall conductivity of visons}
\label{App:hall}
In the presence of Berry curvatures, even non-interacting particles contribute to the (thermal) Hall effect. Independent of the statistics of the particles, bosonic or fermionic, the thermal hall effect at a given temperature $T$ can be calculated from \cite{kappaGeneral}
\begin{align}
	\kappa_{xy}(T) =-\frac{1}{T} \int_0^\infty d\epsilon\, \epsilon^2\, \sigma^v(\epsilon) \frac{\partial n}{\partial \epsilon},\label{eq:kappa}
	\end{align}
	where $n(\epsilon)$ describes the thermal occupation of the particle as function of their energy and $\sigma_{xy}(\epsilon)$ is computed from
\begin{align}
	\sigma_{xy}(\epsilon) =-\sum_\alpha \int \frac{d^2k}{(2\pi)^2} \vec \Omega_{\alpha \vec k}\Theta\left(\epsilon - E^v_{\alpha \vec k}\right)\label{eq:sigmaKappa}
\end{align}
Note that $\sigma_{xy}(\epsilon)$ is in general {\em not} the electrical conductivity at temperature $T$ but is only used to write the formula in a compact way. $\Omega_{\alpha \vec k}$ is the Berry curvature of a band with index $\alpha$. For a single-particle Hamiltonian of the form $\vec H(\vec k) = \vec h(\vec k)\cdot \vec \sigma $ it can be computed from $\vec \Omega_{\alpha \vec k}=\hat{\vec h}\cdot\left(\frac{\partial \hat{\vec h}}{\partial k_x}\cross\frac{\partial \hat{\vec h}}{\partial k_y}\right)$ with the unit vectors $\hat h(\vec k)=h(\vec k)/|h(\vec k)|$. 

To calculate the total thermal Hall effect in the presence of a magnetic field,
we have to compute both the contribution from Majorana fermions and visons. Here we neglect all interaction effects which is only justified in the low-$T$ limit when the density of visons is low.

A magnetic field $h$ induces next-nearest neighbor hopping of Majorana fermions with amplitude $t_{AA}^m$. Such a hopping on the same sublattice, from $A$ to $A$ or $B$ to $B$ sublattice, breaks time-reversal symmetry and opens a gap in the Majorana spectrum. For the calculation of the thermal Hall effect, $t_{AA}^m$ is, however, essential as it renders the Majorana bands topological. The Majorana modes $c_\vec{k}$ and $c_{-\vec{k}}$ can be combined to a complex Fermion, thereby reducing the size of the 1. Brillouin zone (and therefore the integral in Eq.~\eqref{eq:sigma}) by a factor of $2$. The thermal Hall effect is computed from using Eq.~\eqref{eq:kappa} with $n(\epsilon)$ being the Fermi distribution function. 
At low temperature, the Majorana contribution obtains a quantized value
\begin{align}
\kappa_{xy}^m(T) \approx \frac{1}{2} \frac{\pi T}{6} \quad \text{for }\ T \ll t_{AA}^m
\end{align}
In a quantum Hall system one obtains instead  $\kappa_{xy}^m=n \frac{\pi T}{6} $ with integer $n$. The half-integer value of the prefactor $1/2$ arises because
we consider Majorana particles instead of fermions. For larger $T$, when also the upper Majorana band gets occupied, the Majorana contribution drops. Thus it can not explain the peak in $\kappa_{xy}/T$ observed experimentally \cite{kasahara1,kasahara2}.

Exactly the same formalism can be used to calculate also the contribution to the thermal Hall effect arising from visons.  Here we have, however, to take into account that each visons carries a Majorana zero mode. Thus a pair of two visons at large distance from each other carries an extra twofold degeneracy. This gives rise to an extra entropy of $\ln \sqrt{2}=\frac{1}{2} \ln 2$ per vison.
In the low-density limit we can ignore any possible hybridization of these zero modes. Thus we can describe the distribution function in this limit by
\begin{align}
n(E^v_{\alpha,\vec{p}})\approx\exp\!\left(-\frac{E^v_{\alpha,\vec{p}}-T \ln \sqrt{2}}{T}\right)\label{eq:distBoltz}
\end{align}
including the entropic correction due to the zero mode.

The vison single-particle Hamiltonian arising from the field- and $\Gamma$ induced hopping
is given by 
\begin{align}
H^v(\vec p)&=E^{v}_0 \mathbb{1} -\vec h(\vec p)\cdot \vec \sigma \nonumber \\
\vec h(\vec p)&=2 t_h \left(\!\begin{array}{c} \sin(\vec p \cdot \vec \eta_1)\\\cos(\vec p \cdot \vec \eta_2)\\ \sin(\vec p \cdot \vec \eta_3) \end{array}\!\right)+2 t_\Gamma \left(\!\begin{array}{c} \sin(\vec p \cdot (\vec \eta_1+\vec \eta_3))\\\cos(\vec p \cdot (\vec \eta_2+\vec \eta_3))\\ \sin(\vec p \cdot( \vec \eta_2-\vec \eta_1)) \end{array}\!\right)
\end{align}
 with $\eta_1=(\frac{1}{2},\frac{\sqrt 3}{2})$,$\eta_2=(\frac{1}{2},-\frac{\sqrt 3}{2})$ and $\eta_3 = (1,0)$.
The corresponding energies are given by $E^v_{\pm,\vec{p}}=E^v_0 \pm |\vec h(\vec p)|$.

For high temperatures, when the density of visons increases, our approach is not valid any more. The statistics of the visons becomes important and vison-vison and vison-Majorana \cite{Nasu2017} interactions can no longer be ignored.  There will also be skew-scattering of visons and Majorana fermions. Furthermore, the Majorana zero modes start to split when visons approach each other. 
\onecolumngrid\
\begin{center}
	\begin{figure*}[t]
		\includegraphics[width=\textwidth]{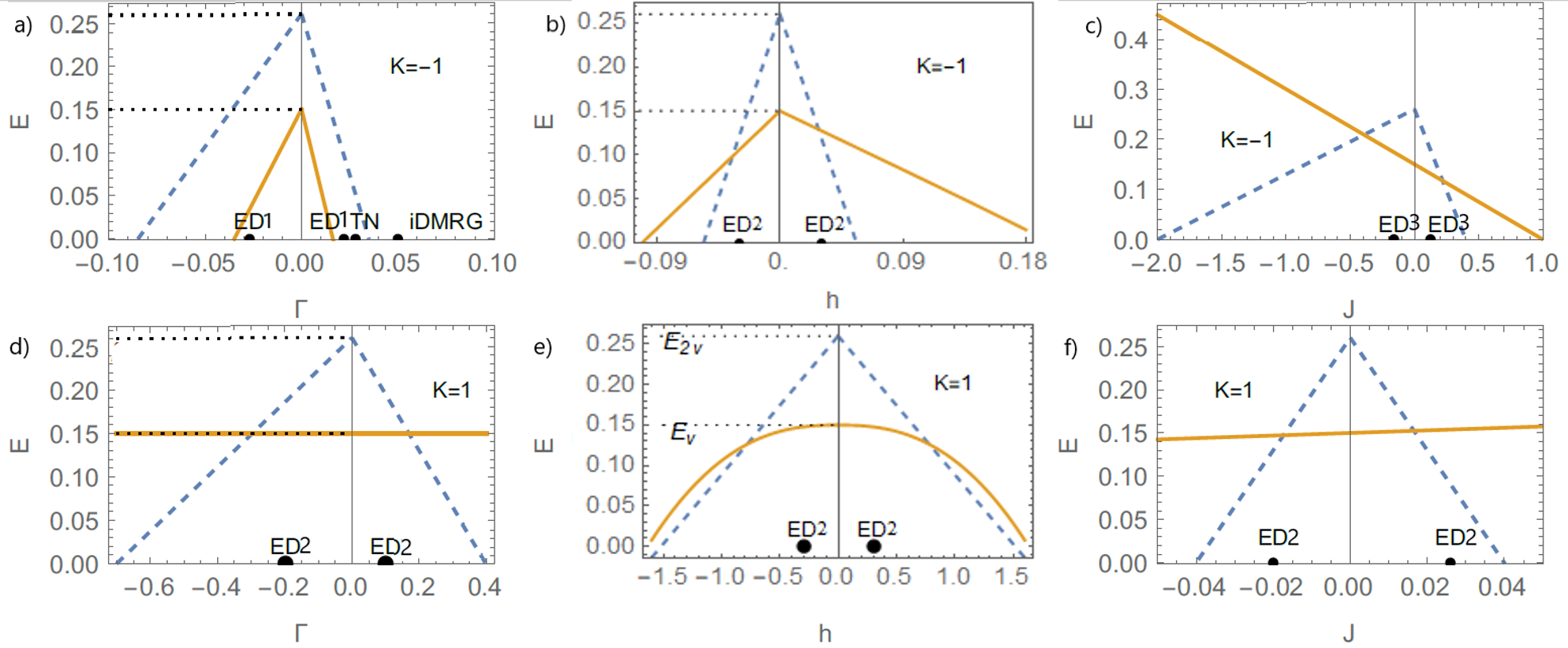}
		\caption{Vison gap as function of $\Gamma$, $h$ and $J$ (solid lines in left, middle, right column) for ferromagnetic (upper panels) and antiferromagnetic (lower panels). In subfigure (e), the vison gap is calculated using the formula \ref{eq:th_afm} which results in a scaling of the form $E_v\propto h^{2.5}$. The dashed line shows the corresponding gap of a vison pair obtained from Zhang {\it et al.}~\cite{Batista}. The thick points show numerical predictions for phase boundaries obtained from the exact diagonalization studies of Ref.~\cite{ciaran} (ED1), Ref.~\cite{gammaRau} (ED2) and Ref.~\cite{ED3} (ED3), from a tensor-network based approach \cite{Tensor} (TN) and from an iDMRG study \cite{gohlkePRR}.\label{fig:compare2}}
	\end{figure*}
\end{center}
\twocolumngrid\
\section{Comparison of vison-pair and single vison gap}
\label{App:compare}

Vison hopping reduces the vison gap and thus is one of several mechanisms which can lead to an instability of the Kitaev spin liquid. Here it is important to consider also a second instability mechanism arising from quasi-bound states of two visons. Formally, such pairs embedded in the Majorana continuum are always unstable and have a finite lifetime.  The tunneling of such vison pairs and their energy  was investigated in an instructive recent study by  Zhang et al. \cite{Batista,zhang}. Note that  vison pairs carry a net flux of zero and thus their properties are very different compared to the single visons studied by us. Furthermore, we also compare the result of the two analytical studies to several numerical studies.
 
In Fig.~9 we show our prediction for the vison gap as function of three different perturbations ($\Gamma$, $h$, and $J$) as solid lines both for the ferromagnetic ($K<0$) and antiferromagnetic ($K>0$) Kitaev model. Furthermore, we show the corresponding predictions of Zhang et al. \cite{Batista} for a vison pair as a dashed line. 
 The analytical treatment breaks down when the vison gap closes but one can use the results to extract trends and leading instabilities.

We first discuss the ferromagnetic Kitaev model, believed to be relevant for materials like $\alpha$-RuCl$_3$ \cite{FMKitaev1,FMKitaev2,FMKitaev3}. 
When perturbed by a $\Gamma$ term, our results suggest that the leading instability arises from the closing of the single-vison gap, see Fig.~9a. Linear order perturbation theory obtains a closing of the gap at values roughly consistent with exact diagonalization (ED) results \cite{ciaran,gammaRau} and a tensor network calculation \cite{Tensor}. Note, however, that a recent iDMRG study \cite{gohlkePRR} predicts an increased stability of the spin liquid phase.

The situation is very different when one considers perturbations by a magnetic field $h$ shown in Fig.~9b. Already for rather small fields,  vison pairs have a lower energy compared to single visons suggesting that the condensation of vison pairs (or more complicated objects) is a prime candidate for the instability. The predicted location of the transition is again roughly consistent with ED studies.

For a perturbation by $J$, we do not predict any vison motion to linear order in $J$ but there is a  trivial change of the vison gap when one absorbs part of the Heisenberg coupling in the Kitaev coupling, $K \to K+J$. Here linear order perturbation theory suggests again that vison pairs become gapless first. In this case, however, the ED calculation predicts that the spin liquid is unstable for very small values of $J$. Therefore most likely other types of excitations or more complex bound states \cite{Batista} may drive the transition.

In the antiferromagnetic case, $K>0$, shown in the lower panel of Fig.~\ref{fig:compare2} our theory makes no direct prediction for $\Gamma$ and $J$ perturbations as there is no vison hopping to linear order. For the $h$ perturbation, we find that the single vison gap closes at a similar critical field as the vison pair. Although the bare vison pair gap closes at a large field value, well beyond the perturbative limit, Ref.\cite{Batista} also reported a smaller critical field $\approx 0.5 K$ where a transition to a different spin liquid phase happens due to the interplay of hybridisation of the vison pairs and Majorna fermions and their dynamics. Compared to the ferromagnetic case, the ED results show that the system is much more stable with respect to perturbations by $\Gamma$ and $h$, roughly consistent with the absence of single-vison tunneling linear in $\Gamma$ or $h$ in this case. The high sensitivity of the spin liquid towards tiny values of $J$, Fig.~\ref{fig:compare2} f, is, most likely, connected to the tunneling of vison pairs  \cite{Batista}.

\bibliography{references}
\end{document}